\documentclass[twocolumn]{aa}  

\usepackage{graphicx}
\usepackage{txfonts}
\usepackage{xspace}
\usepackage{colortbl}
\usepackage{xcolor}
\usepackage{amsmath}
\usepackage{soul}

\newcommand{\kms}{$\rm km \, s^{-1}$\xspace}
\newcommand{\ohhdp}{o$\rm H_2D^+$\xspace}
\newcommand{\hhdp}{$\rm H_2D^+$\xspace}
\newcommand{\xe}{$x(\mathrm{e^-})$\xspace}
\newcommand{\ohdp}{\mathrm{oH_2D^+}}
\newcommand{\phdp}{\mathrm{pH_2D^+}}
\newcommand{\pdhp}{\mathrm{pD_2H^+}}
\newcommand{\odhp}{\mathrm{oD_2H^+}}
\newcommand{\hp}{\mathrm{pH_3^+}}
\newcommand{\ho}{\mathrm{oH_3^+}}
\newcommand{\cdo}{$\rm C^{18}O$\xspace}
\newcommand{\htcop}{$\rm H^{13}CO^+$\xspace}
\newcommand{\dcop}{$\rm DCO^+$\xspace}
\newcommand{\hcop}{$\rm HCO^+$\xspace}
\newcommand{\tex}{$T_\mathrm{ex}$\xspace}
\newcommand{\olineh}{o$\rm H_2D^+(1_{1,0} - 1_{1,1})$\xspace}
\newcommand{\crir}{$\zeta_2$\xspace}
\newcommand{\crirs}{$\zeta_2\rm ^s$\xspace}

\definecolor{orange}{rgb}{1,0.5,0}
\definecolor{sred}{rgb}{.5,0,0}

\begin{document}

   \title{Testing analytical methods to derive the cosmic-ray ionisation rate in cold regions via synthetic observations}


   \author{E. Redaelli
          \inst{1}
          \and
          S. Bovino\inst{2,3,4}
          \and
          A. Lupi \inst{5,6}
          \and
          T. Grassi \inst{1}
          \and
          D. Gaete-Espinoza \inst{2}
          \and
          G. Sabatini \inst{7,3}
          \and
          P. Caselli \inst{1}
          }

   \institute{Centre for Astrochemical Studies, Max-Planck-Institut f\"ur extraterrestrische Physik, Gie{\ss}enbachstra{\ss}e 1, 85749 Garching bei M\"unchen, Germany\\
              \email{eredaelli@mpe.mpg.de}
         \and
          Departamento de Astronom\'ia, Facultad Ciencias F\'isicas y Matem\'aticas, Universidad de Concepci\'on, Av. Esteban Iturra s/n Barrio 
Universitario, Casilla 160, Concepci\'on, Chile
            \and 
     INAF, Istituto di Radioastronomia --- Italian node of the ALMA Regional Centre (It-ARC), Via Gobetti 101, 40129 Bologna, Italy     
         \and 
    Dipartimento di Chimica, Universit\`a degli Studi di Roma ``La Sapienza'', P.le Aldo Moro 5, 00185 Roma, Italy
         \and
         Dipartimento di Scienza e Alta Tecnologia, Universit\`a degli Studi dell'Insubria, via Valleggio 11, I-22100 Como, Italy  \and
         INFN, Sezione di Milano-Bicocca, Piazza della Scienza 3, 20126 Milano, Italy
         \and
         INAF, Osservatorio Astrofisico di Arcetri, Largo E. Fermi 5, I-50125, Firenze, Italy
     }

   \date{xxx}

 
  \abstract
   {Cosmic rays (CRs) heavily impact the chemistry and physics of cold and dense star-forming regions. However, characterising their ionisation rate is still challenging from an observational point of view.}
   {In the past, a few analytical formulas have been proposed to infer the cosmic-ray ionization rate \crir from molecular line observations. These have been derived from the chemical kinetics of the involved species, but they have not been validated using synthetic data processed with a standard observative pipeline. We aim to bridge this gap.}
   {We perform the radiative transfer on a set of three-dimensional magneto-hydrodynamical simulations of prestellar cores, exploring different initial \crir, evolutionary stages, types of radiative transfer (e.g. assuming local-thermodynamic-equilibrium conditions), and telescope responses. We then compute the column densities of the involved tracers to determine \crir{}, using, in particular, the equation proposed by \cite{Bovino20} {and by \cite{Caselli98}, both used nowadays}.}
   {Our results confirm that the method of \cite{Bovino20} accurately retrieves the actual \crir within a factor of $2-3$, in the physical conditions explored in our tests. Since we also explore a non-local thermodynamic equilibrium radiative transfer, this work indirectly offers insights into the excitation temperatures of common transitions at moderate volume densities ($n\approx 10^5 \, \rm cm^{-3}$). We have also performed a few tests using the formula proposed by \cite{Caselli98}, which overestimates the actual \crir by at least two orders of magnitudes. We also consider a new derivation of this method, which, however, still leads to large overestimates.}
   {The method proposed by \cite{Bovino20}, further validated in this work,  
   represents a reliable method to estimate \crir in cold and dense gas. We also confirm that the former method by \cite{Caselli98}, as already pointed out by its authors, has no global domain of application, and should be employed with caution.}


   \maketitle
%

\section{Introduction}

Cosmic rays (CRs) are energetic, ionised particles found ubiquitously in the interstellar medium (ISM). In the densest gas phases, they play a key role in triggering the chemistry and regulating the thermodynamical balance. In molecular gas, CRs collide with H$_2$ molecules, most of the times ionising them via the reaction
\begin{equation*}
 \rm  CR+ \mathrm{H_2} \longrightarrow \mathrm{H_2^+} + \mathrm{e^-} + CR \; ,
\end{equation*}
after which the highly reactive H$_2^+$ immediately reacts with another H$_2$ molecule, producing the trihydrogen cation H$_3^+$. By doing so, CRs set the ionisation state of dense matter, where the photodissociating UV field is completely attenuated. This, in turn, strongly affects the dynamics of the dense gas. The electron fraction \xe determines the degree of coupling between the magnetic fields and the matter, and it is, therefore, linked to the time scale for ambipolar diffusion. This corresponds to the drift between the neutral and the ionised flows, which is one of the proposed mechanisms to dissipate magnetic flux and to allow the gravitational collapse \citep{Mouschovias76}. \par
 H$_3^+$  is a pivotal species also for the ISM chemical evolution since it drives the rich ion chemistry, which is ultimately responsible for the production, for instance, of CO and gas-phase water in cold regions. Furthermore, by reacting with deuterated hydrogen HD,  H$_3^+$ is converted into \hhdp, which is the starting point of the deuteration process in the gas phase \citep[see e.g.][and references therein]{Ceccarelli14}.
\par
CRs are crucial for the physics and chemistry of star-forming regions. The key question is how to measure their effects, particularly their ionisation rate per hydrogen molecule (\crir). How to infer \crir from typical observables has been the focus of several works, dating back to the '70s \citep[see e.g.][]{Guelin77, Wootten79}. In the diffuse medium ($A_\mathrm{V}\lesssim 1 \, \rm mag$), H$_3^+$ can be directly observed in absorption towards bright infrared sources, and its relatively simple kinetics can be solved to infer \crir. This is the approach followed for instance by \cite{McCall03,Indriolo07, Indriolo12}, with typical results of the order of $\text{\crir} \approx 10^{-16} \, \rm s^{-1}$. \par
At higher densities, the absorption lines of H$_3^+$ cannot be detected anymore. In this situation, \cite{Guelin77, Guelin82} proposed an analytic expression containing \crir, based on the kinetics of \dcop{} and \hcop{}. This work was later expanded by \citet[][hereafter CWT98]{Caselli98}, who developed a system of two equations to infer first the ionisation fraction \xe, and then \crir. Those authors ultimately used a more comprehensive chemical code to infer these parameters in a sample of dense cloud cores (see their Sect. 5.1 and Table 7), resulting in \crir values on average lower than with the analytic expression. However, the analytic equations are still used nowadays (e.g. \citealt{Cabedo22}), due also to the fact that they depend on common observable quantities, such as\footnote{$X(\rm A)$ denotes the abundance of species A with respect to molecular hydrogen.} $R_\mathrm{H} = \rm X(HCO^+)/ X(CO)$ and the deuteration level of \hcop, $R_\mathrm{D} = \rm X(DCO^+)/ X(HCO^+)$. It is important to notice, however, that this approach was based on several assumptions, such as the abundance of HD. Furthermore, it was derived in specific conditions (e.g. at a temperature of $T_\mathrm{K} = 10\, \rm K$), which prevents its generalisation.  
\par
More recently, \citet[][hereafter BFL20]{Bovino20} suggested a new analytical approach to determine \crir in cold gas, based on the detection of \hhdp (and, in particular, of its ortho spin state, \ohhdp, which can be observed from the ground). This method is, in turn, based on the formulation of \cite{Oka19}, and the underlying idea is to infer the abundance of H$_3^+$ from that of its deuterated forms, together with the deuterium fraction measured from \hcop isotopologues. It relies on fewer assumptions than CWT98, but it requires \ohhdp data, which are observationally more expensive in terms of observing time. \cite{Sabatini20} applied this implementation to a sample of high-mass star-forming clumps, using observations made with the Atacama Pathfinder EXperiment 12-meter submillimeter telescope (APEX; Gusten et al. 2006) and the Institut de Radioastronomie Millimétrique (IRAM) 30m single-dish telescope. Their analysis yielded $7\times 10^{-18} < \zeta_2 / \rm s^{-1} < 6\times 10^{-17}$, in line with theoretical predictions \citep{Padovani18}. \cite{Sabatini23} used the BFL20 equation to estimate \crir maps at high resolution in two high-mass clumps, using data from the Atacama Large Millimeter and sub-millimeter Array (ALMA), finding a remarkable agreement with the most recent predictions of cosmic-ray propagation and attenuation \citep{Padovani22}.
\par
Another possible approach is based on having an underlying chemical model to interpret the observational results, possibly using multiple tracers. This can be done either by comparing the abundances and column densities obtained from observations to the results of chemical codes 
\citep[as done, for instance, by][]{Caselli98, Ceccarelli14, Fontani17, Favre18}, or employing radiative transfer analysis. In the latter case, used for instance in \cite{Redaelli21b}, one compares the synthetic spectra of several transitions based on the molecular abundances from the chemical modelling to the observed lines. This approach is more sophisticated, and can potentially constrain \crir more accurately, but it also depends on the employed model and the detail of the physics included to study the given source. However, its applicability is limited by the considerable combined runtimes of a chemical code and radiative transfer.
\par
When evaluating \crir is required in large surveys, an analytical expression presents a clear advantage in terms of applicability and use of computational resources. The equation by BFL20 was tested on several setups of physicochemical three-dimensional simulations varying, for instance, the collapse speed, or the initial H$_2$ ortho-to-para ratio (OPR). However, a validation based on simulated observations was never performed. This is crucial to test if (and how) the results are affected by observational effects depending, for instance, on the selected transitions or the line opacities. The same is true for the CWT98, which moreover was never tested, either on a simulation set or on synthetic observations. In this paper, we aim to i) investigate the effects of telescope response and of several possible observational biases and ii) compare two widely used equations to infer \crir from observable tracers, and look for the potential applicability limits of the formulas. For these reasons, we have selected two simulation setups of a dense core, where the \crir is known by construction, and it varies in the typical range predicted by models at high densities {($\approx {10^{-18}}-10^{-16}\rm \, s^{-1}$)}. These simulations are used as input for a radiative transfer code, producing synthetic spectra of the chemical tracers involved. These are then post-processed to simulate either single-dish or interferometric-like observations. From this point, the analysis follows the standard procedure to apply the analytical expressions: the datacubes are converted into column densities and abundances maps, on which the formulas can be evaluated. The resulting \crir maps are compared to the actual input values of the ionization rate. 
\par
The paper is organised as follows: Sect.~\ref{sec:ana_approaches} introduces the analytical methods put to the test. Sect.~\ref{sec:sim_postproc} illustrates the simulation setup, the radiative transfer analysis, and the post-processing procedure to infer column density maps. The \crir maps are computed and described in Sect.~\ref{sec:results}. First, we focus on the BFL20 formulation, performing several runs starting from a reference one (where $\zeta_2 = 2.5\times 10^{-17}$ s$^{-1}$ and the evolutionary time is $t = 100 \, \rm kyr$), and then exploring distinct post-processing types, radiative transfer methods, and a set of rotational lines at different evolutionary stages (Sect.~\ref{sec:results_bovino}). In Sect.~\ref{sec:analysis_caselli98}, instead, we test the CWT98 analytical approach. Sect.~\ref{sec:realObs} presents the comparison of the two methods on literature values of real observed objects. The results are discussed in Sect.~\ref{sec:discussion}. A final summary with concluding remarks is presented in Sect.~\ref{sec:summary}.

\section{Retriving \crir{} using analytical expressions\label{sec:ana_approaches}}
The first analytical expression we aim to test is the one proposed by BFL20. In Appendix~\ref{app:Ste} we present the detailed derivation of the final equation based on observable quantities (i.e. molecular column densities $N_\mathrm{col}$), as
    \begin{equation}
    \zeta_2 = k_\mathrm{CO}^\mathrm{oH_3^+} \frac{1}{3}\times X(\mathrm{CO}) \times \frac{N_\mathrm{col}(\mathrm{oH_2D^+})}{R_\mathrm{D}}\frac{1}{L}    \; ,
    \label{eq:crir_observables}
    \end{equation}
where the deuterium fraction of \hcop is
\begin{equation*}
    R_\mathrm{D} = \frac{N_\mathrm{col}(\text{\dcop})}{68 \times N_\mathrm{col}(\text{\htcop})} \; , 
\end{equation*}
and the CO abundance (with respect to H$_2$) is
\begin{equation*}
X(\mathrm{CO})  = \frac{557 \times N_\mathrm{col}(\text{\cdo})}{ N(\mathrm{H_2})}  \; . 
\end{equation*}
In Eq.~\eqref{eq:crir_observables}, $k_\mathrm{CO}^\mathrm{oH_3^+}$ is the destruction rate of oH$_3^+$ by CO (see reaction rate $k_7$ in Appendix~\ref{app:Ste}), assumed here to be the main destruction path for H$_3^+$, which holds when the CO depletion factor is up to $\approx 100$ assuming $x(\mathrm{e^-}) \approx 10^{-8}$ (see also the discussion in Appendix~\ref{app:Ste} for more details). $L$ is the path length over which the column densities are estimated. Note that for tracers that are typically optically thick (CO and \hcop), we use the corresponding optically thin isotopologues, assuming standard isotopic ratios for the local ISM: $\rm ^{18}O/^{16}O = 557$ and $\rm ^{13}C/^{12}C = 68$ \citep{Wilson99}. This choice is consistent with that made later in the radiative transfer analysis (see  Sect.~\ref{sec:polaris}). We stress that Eq.~\eqref{eq:crir_observables} can be used only until when \hhdp is the dominant deuterated species of H$_3^+$, i.e. at the early stages of star-forming regions.
\par
We aim to test the performance of the approach proposed by CWT98 as well, exploring its applicability and comparing it with the BFL20 approach. In these regards, in Sect.~\ref{sec:analysis_caselli98} we first use their equations 3 and 4, as they are. This is motivated by the fact that some papers have been using them in that exact formulation (although CWT98 already employed chemical modelling to interpret the observational results). The numerical constants in those equations, however, were derived at a temperature of $10\, \rm K$, and without including ortho- and para-states of the involved species. To update the work of CWT98, in Appendix~\ref{app:Paola} we follow the same approach but include the spin-state separation of H$_2$, H$_3^+$, and \hhdp. We also update the reaction rates involved with the most recent values, and we keep their temperature dependency.

\section{Simulations and post-processing\label{sec:sim_postproc}}
In this work, we post-process three-dimensional simulations of prestellar cores to produce the observables needed to test the aforementioned analytical methods. In the following subsections, we describe the set of simulations used in our test, the details of the radiative transfer, and the type of post-processing performed afterwards. Finally, we discuss how we recover the molecular column densities and the gas total column density from the synthetic observations. 
\subsection{The simulation setup\label{simulations}}
We use a set of three-dimensional magneto-hydrodynamical (MHD) simulations of prestellar cores, with constant and variable cosmic-ray ionisation rates, obtained with the code \textsc{gizmo} \citep{Hopkins2015}. The simulations evolve an isothermal, turbulent and magnetised Bonnor-Ebert sphere of 20 M$_\odot$ with a radius of 0.17$\,$pc, resembling a collapsing prestellar core. After 100$\,$kyr of evolution, we obtain a low-mass object, with a total mass of $\sim 3 \, \rm M_\odot$ in the central $10000 \, \rm AU$ (corresponding to an average density of $\langle n( \rm H_2) \rangle \sim 10^5 \, cm^{-3}$). The gas and dust temperatures are $T_\mathrm{k} = T_\mathrm{dust} = 15 \, \rm K$. The size of the total simulation box is 0.6$\,$pc. For this work, we focus on the central 0.3$\,$pc containing the Bonnor-Ebert sphere.\par
{The simulations include a state-of-the-art deuterated and spin-state chemical network, advanced in time alongside hydrodynamics. We refer to \cite{Bovino19} for the complete description of the physical and chemical initial conditions. We highlight that the chemical code includes molecular depletion and thermal and cosmic-ray-induced desorption, but no further surface chemistry is considered. This assumption is justified since, at the low temperatures considered, the thermal desorption of any species is negligible, and the cosmic-ray-induced desorption timescale is longer than the collapse one \citep[cf.][]{Bovino19}. Hence, surface chemistry would have negligible effects on the final abundances of gas phase species. The initial H$_2$ ortho-to-para ratio is $\rm OPR=0.1$, consistent with the values obtained by large-scale simulations \citep{Lupi21}.} \par
To investigate different ionisation states, we first use a simulation with constant $\zeta_2^\mathrm{s} = 2.5\times 10^{-17}$ s$^{-1}$, a value typically assumed for dense regions (this corresponds to the M1 case of \citealt{Bovino19}). Throughout this paper, we use the superscript "$\rm s$" to denote the input \crirs of the simulations, to avoid confusion with the \crir retrieved from the synthetic observations. {We perform an additional test on a simulation performed with constant $\text{\crirs}= 2.5\times 10^{-18} \, \rm s^{-1}$, to expand the range of input \crir explored and to be consistent with the tests performed by \cite{Bovino20}}. We also simulate a variable \crirs model, by post-processing the same simulation by employing the framework developed by \cite{Ferrada-Chamorro21} where \crir is varied according to the density-dependent function reported in \cite{Ivlev19}. The details of the latter simulations will be presented in a forthcoming paper (Gaete-Espinoza et al., in prep). This set of simulations hence covers the typical \crir values predicted by the most recent models of CR propagation at high densities \citep[for $N \rm (H_2) \gtrsim 10^{22} \, cm^{-2}$ $\zeta_2 \lesssim \text{a few} \times 10^{-16}\, \rm s^{-1}$; cf.][]{Padovani22}.

\subsection{Description of tested runs \label{sec:testedRuns}}
Since our goal is to compare the results of analytic expressions to infer \crir from observables, we test a variety of combinations of types of radiative transfer (assuming or deviating from the local-thermodynamic-equilibrium), of rotational transitions (ground state lines, or higher J transitions), and of post-processing (simulating single-dish or interferometric observations). Starting from a reference run, we modify one parameter at a time, to evaluate its effects. We perform a total of eight tests, reported in Table~\ref{tab:tests} and described in the following text. The details of how the radiative transfer is performed are discussed in Sect.~\ref{sec:polaris}, whilst Sect.~\ref{sec:postproc} describes how the telescope response is simulated.
\par
Runs~1 and 2 (the latter is considered the reference one throughout the paper) use the simulation with constant $\zeta_2^\mathrm{s} = 2.5\times 10^{-17}\rm \, s^{-1}$ at an evolutionary time of 50 and $100\,$kyr, respectively. The radiative transfer is performed in local-thermodynamic-equilibrium approximation (LTE), focussing on the molecular lines in the $215-370\, \rm GHz$ range. The post-processing is single-dish-like, with a final beam size of $27''$. {Run~3 tests the low-\crirs case, where the only difference with respect to the reference run is the value $\zeta_2^\mathrm{s} ={2.5\times 10^{-18}\rm \, s^{-1}} $.
Run~4 explores the interferometric-like post-processing,} (see Sect.~\ref{alma_analysis}). In runs~5 and 6 we modify the kind of radiative transfer, using a non-LTE approach. In particular, in run 5 we simulate again the high-J transition of \cdo, \dcop, and \htcop. In run 6, instead, we aim to explore the effect of targeting the (1-0) transitions of \cdo, \htcop, and \dcop. This is beneficial to the observational studies that trace the $3 \, \rm mm$ lines of these molecules, as in the pioneering work of \cite{Caselli98}. We also adopt a large beam size of $\approx 70''$, to simulate poorly resolved observations, where the beam area is comparable to the source size. The final two runs adopt the LTE analysis and single-dish-like post-processing, performed on the simulation with variable \crirs at 50$\,$kyr (run 7) and 100$\,$kyr (run 8), respectively.

\begin{table*}
    \centering
    \renewcommand{\arraystretch}{1.4}
    \caption{Properties of the simulations performed in this work. Run~2, highlighted with the asterisk symbol, is the reference one.}\label{tab:tests}
    \begin{tabular}{cccccccc}
    \hline
     Run & Time & \crirs & Radiative tran.\tablefootmark{a} & Lines $\nu$\tablefootmark{b} & Post-processing type\\
    \hline
    \hline
  1 & $50\, \rm kyr$ &  $2.5\times 10^{-17}\rm \, s^{-1}$ &LTE & $215-230\,\rm GHz$ & single-dish \\
  2$^*$ & $100\, \rm kyr$ &  $2.5\times 10^{-17}\rm \, s^{-1}$ &LTE & $215-230\,\rm GHz$ & single-dish \\
 3 &  100$\, \rm$ kyr &  ${2.5\times 10^{-18}\rm \, }$ s$^{-1}$ &  LTE & ${215-230}\,\rm $GHz &  single-dish \\
  4 & $100\, \rm kyr$ &  $2.5\times 10^{-17}\rm \, s^{-1}$ &LTE & $215-230\,\rm GHz$ & ALMA-like \\
5 & $100\, \rm kyr$ &  $2.5\times 10^{-17}\rm \, s^{-1}$ &LVG & $215-230\,\rm GHz$ & single-dish \\
  6 & $100\, \rm kyr$ &  $2.5\times 10^{-17}\rm \, s^{-1}$ &LVG &$72-110\,\rm GHz$ & single-dish \\
  7 & $50\, \rm kyr$ & variable &LTE & $215-230\,\rm GHz$ & single-dish \\
  8 & $100\, \rm kyr$ & variable  &LTE & $215-230\,\rm GHz$ & single-dish \\
\hline
  \end{tabular}
  \tablefoot{
\tablefoottext{a}{Type of radiative transfer used to produce the synthetic observations.}\\
\tablefoottext{b}{Frequency coverage of the simulated molecular lines. $215-230\,\rm GHz$ indicates we use the \dcop and \htcop (3-2) and the \cdo (2-1) lines, whilst $72-110\,\rm GHz$ refers to the run using the lowest-J transitions. } 
}
\end{table*}

\subsection{Radiative transfer of MHD simulations\label{sec:polaris}}
The total gas column density distribution and the molecular column densities are involved in the equations to infer \crir (see Sect.~\ref{sec:ana_approaches}). The radiative transfer of the dust and the molecular lines is performed with the \textsc{polaris} code\footnote{Latest version available at \url{https://github.com/polaris-MCRT/POLARIS}. For this work, we used a custom version, where we corrected an issue in the conversion between mass fractions and number densities.} \citep{Reissl16, Brauer17}.
Concerning the radiative transfer of the dust, we set the gas mean molecular weight to $\mu = 2.4$ \citep{Kauffmann08} and the gas-to-dust mass ratio to $100$ \citep{Hildebrand83}. We simulate the dust emission at a wavelength of $870 \, \rm \mu m$ (corresponding to $345\, \rm GHz$). From an observational perspective, this was the wavelength of the LABOCA instrument mounted on  APEX, which was used to perform the all-sky survey ATLASGAL \citep{Schuller09}. It is also close to the James Clerk Maxwell Telescope (JCMT) SCUBA II longer wavelength ($850 \rm \, \mu m$). Finally, it represents the standard frequency for ALMA continuum observations in Band 7. We stress, however, how the choice of wavelength does not impact the results. For the dust model, we assume pure silicate grains, with opacities taken from \cite{LaorDraine93}\footnote{These are listed in the file silicate\_ld93.nk, available in the \textsc{polaris} package.}. The grain size distribution is a standard MRN \citep{Mathis77}, with a power-law index of $-3.5$ between $5\rm \, nm$ and $0.25 \rm \, \mu m$. The grain density is $3.5\, \rm g \, cm^{-3}$, which is consistent with the value assumed in the simulations. These parameters are likely different from the real dust populations within dense cores, where for instance a mixture of carbonaceous and silicate grains is expected. However, our goal is not to reproduce the exact properties of a real dust population, but to be consistent in the various steps of the analysis, from the simulation to the radiative transfer.  \par
{For the molecular tracers, \textsc{polaris} needs as input the spectroscopic description of the simulated transitions, which are summarised in Table~\ref{tab:transitions}.} In the case of \ohhdp, the only line accessible from the ground is the $(1_{1,0} - 1_{1,1})$ one at $372 \, \rm GHz$ \citep{Caselli03}. This can be observed, for instance, by ALMA in Band 7 and by APEX using, e.g. SEPIA345 or LAsMA. Concerning the other tracers (\cdo, \dcop, and \htcop), their transitions at $215-260 \, \rm GHz$ are commonly observed. {However, several studies focus on their lower-J transitions at $3\,$mm, and hence we test also these lines in run 6.} All molecular transitions are simulated over a total velocity range of $7.5\,$\kms and a velocity resolution of $0.1\, $\kms. We assume that the local standard-of-rest velocity of the source is $0\,$\kms. 
\par
\textsc{polaris} can perform different approaches of radiative transfer, including LTE and large velocity gradient (LVG). We perform six runs assuming LTE conditions, which allow us to focus initially on the effect of the radiative transfer itself and of the response of the telescope on the inferred \crir values. However, lines with high critical densities ($n_\mathrm{c} \approx 10^5 -10^6 \, \rm cm^{-3}$), such as the high-J transitions of \htcop and \dcop, and \ohhdp, are likely to be sub-thermally excited. Two runs (n.~5 and 6) hence explore a more realistic case, using the LVG approach.
\par
\textsc{polaris} accepts a variety of grid types as input, in particular Voronoi grids. The simulations we consider are obtained with \textsc{gizmo}, which samples the fluid using a set of discrete tracers representing a sort of cells with smoothed boundaries. In this respect, converting this volume discretisation to a Voronoi tessellation is the most natural and consistent choice, despite some differences existing between the two volume partition schemes\footnote{As a consistency check, we compared the cell volume obtained in the simulation with that of the corresponding Voronoi cell, finding negligible differences, and only for cells with very asymmetric shapes.}. The outputs of the simulations are hence prepared in the form of a Voronoi grid. In order to properly treat boundary cells, we added at the edges of the region of interest a set of virtual particles placed according to a cubic regular grid. Virtual particles are placed at 1.5 times the simulated region size to avoid artefacts and passed to the \textsc{SciPy} package \citep{Virtanen20} to build the Voronoi cells. The grid is then cut to match the original region, and the information associated with every cell is passed to \textsc{polaris}, including the IDs of the cell and its neighbours. In particular, the gas density, gas and dust temperatures, velocity field (three dimensional), and the molecular mass fraction for each species ($f_\mathrm{mol} = \rho_\mathrm{mol}/\rho_\mathrm{tot}$) are the input of the radiative transfer. Note that the chemical code does not treat oxygen or carbon fractionation. The mass fractions of \cdo and \htcop are hence derived from the main isotopologues' ones, using the same standard isotopic ratios assumed in Sect.~\ref{sec:ana_approaches}). \par
{In all radiative transfer analyses}, the grid size of the output maps or cubes produced by \textsc{polaris} is set to 256 pix $\times$ 256 pix. We aim to produce synthetic observations both for a single dish-like and for an interferometer-like case, with the distance of the source set to $170\, \rm pc$ and $2\,$kpc, respectively\footnote{The former value is within $30\, \rm pc$ from the distance of nearby low-mass star-forming regions, such as parts of Taurus, the Pipe, and Lupus \citep{Dzib18, Galli19}. The larger distance, instead, is consistent with that of some of the closest infrared-dark clouds, see e.g. \cite{Sanhueza19}.}. The final pixel and field-of-views (FoV) are $1''.4$ and $6'\times 6'$ (single-dish), and $0''.12$ and $30''\times 30''$ (interferometer-like).

 \begin{table*}
    \centering
    \renewcommand{\arraystretch}{1.4}
    \caption{Properties of the molecular line transitions used for this work. }\label{tab:transitions}
    \begin{tabular}{ccccccccc}
    \hline
     Species & Transition & $\nu$ &  ALMA band\tablefootmark{a} & Single-dish\tablefootmark{b} & $g_\mathrm{u}$ & $A_\mathrm{ul}$& $E_\mathrm{u} / k_\mathrm{B}$ & $T_\mathrm{ex}$\tablefootmark{c}\\
            &       &(GHz) &            &            &                  &  ($\rm s^{-1}$)   &  (K)   &  (K)   \\     
    \hline
    \hline
	 \cdo & 1-0 & 109.78 & 3  & IRAM30m & 3 & $6.27\times 10^{-8}$ & $5.27$ & 15\\
	        & 2-1 & 219.56 & 6  & APEX & 5& $6.01 \times 10^{-7}$ & $15.8$ & 15\\
	\htcop   & 1-0 & 86.754& 3   & IRAM30m & 3 & $3.85\times 10^{-5}$ &$4.16$ & 10\\
	        & 3-2 & 260.26 & 6   & APEX & 7 &$1.34\times 10^{-3}$ & $25.0$& {5.5} \\
     \dcop & 1-0 & 72.039 & -   & IRAM30m & 3 &$2.21\times 10^{-5}$ & $3.46$ & 10\\
            & 3-2 & 216.11 & 6   & APEX &7 &$7.66\times 10^{-4}$ &$20.7 $ & {5.5}\\
     \ohhdp & $1_{1,0} - 1_{1,1}$ & 372.42 &  7   & APEX & 5&$1.10\times 10^{-4}$ &  $17.9$ & 10\\
    \hline
  \end{tabular}
\tablefoot{
\tablefoottext{a}{ALMA band that covers the line frequency. Note that the \dcop (1-0) transitions cannot be covered by any ALMA receiver.}\\
\tablefoottext{b}{Examples of single-dish facilities that can detect the line. } \\
\tablefoottext{c}{Excitation temperature values used in the case of LVG radiative transfer.}
}
\end{table*}

\subsection{Post-processing of the \textsc{polaris} output \label{sec:postproc}}
The output of \textsc{polaris} consists of bi-dimensional maps (in $\rm Jy \, pix^{-1}$), one for each wavelength for the continuum emission (a single one at $870\, \rm \mu m$ in our case) or one for each velocity channel set for the molecular lines. In the latter case, the first stage is to build the position-position-velocity datacube concatenating all the velocity channels. We now describe the approaches used to simulate a single-dish-like or interferometric response.
\subsubsection{Single-dish analysis}
In this case, we convolve the continuum maps and the molecular line datacubes to a specific beam size. For all the tests performed with the higher J transitions, we chose a beam size of $27''$. This corresponds approximately to the APEX beam size at the lowest frequency in the sample, \dcop (3-2) at $216\,\rm GHz$. In the case of run 6, where we simulate the lower $J=(1-0)$ lines (see Sect.~\ref{sec:run5_6} for more details), we select a beam size of $70''$. It is aimed at determining the effects of poorly resolved observations. \par
We 
introduce pixel by pixel in the data cubes and in the continuum fluxes some artificial Gaussian noise with zero mean and $rms$ standard deviation. For the continuum maps, we use $rms = 15 \, \rm mJy \, beam^{-1}$ and $rms = 100 \, \rm mJy \, beam^{-1}$ for the cases at $27''$ and $70''$ of resolution, respectively. Concerning the line datacubes, we inject a noise with $rms=100\, \rm mK$.  For \ohhdp in run 1, and all lines in run 5, this sensitivity is insufficient for significant detections (see Sect.~\ref{sec:run1_2} and \ref{sec:run5_6} for more details). In these runs, we reduce the noise level to  $rms=50\, \rm mK$. These values are consistent with the typical $rms$ of observational campaigns with APEX \citep[cf.][]{Sabatini20}. {Run~3, performed with the lowest \crirs value, present faint lines, and the noise level is reduced to $rms=1\, \rm mK$  (cf. Sect.~\ref{sec:run3}).}

\subsubsection{Interferometer-like analysis \label{alma_analysis}}
To simulate interferometer-like observations, we focus on ALMA, which can cover the frequencies of the transitions analysed here, except for \dcop (1-0). We hence use the output of \textsc{polaris} as input for the task \textsc{simobserve} of \textsc{casa} (version 6.4.3). Due to current limitations of \textsc{simobserve}, it is not possible to add total power at the desired sensitivity. We hence simulate only the 12m and 7m-array observations. We chose the Cycle 8 configuration sets. The integration times are selected using the corresponding Observing Tool (OT), setting the requested angular resolution to $1''$ and the desired noise level to $100\, \rm mK$ 
and to $25 \, \mu \rm Jy$ for the continuum simulations. Table \ref{tab:ALMA-like} summarises the integration times used in each run of \textsc{simobserve}. Concerning the noise, we let the task construct the atmospheric model (using the option \textsc{thermal\_noise = tsys-atm}).
\par
The task \textsc{simobserve} is called separately to simulate the 12m and the 7m array observations. The output visibilities are hence combined (using \textsc{concat}), making sure that the relative weights are correct\footnote{Following the instructions at \url{https://casaguides.nrao.edu/index.php/DataWeightsAndCombination}.}. After that, the concatenated visibilities are imaged using the \textsc{tclean} task. We use the \textsc{multiscale} deconvolver (scales: $[0,5,15]\times$pixel size), which is an appropriate choice in case of extended emission, such as in the simulated data. We select a \textsc{briggs} weighting, with \textsc{robust = 0.5}. The noise threshold is set to $2\sigma$. {The final datacubes (or 2D images for continuum observations) have a FoV of $30'' \times30''$, sampled with $250 \rm \, pix \times 250\, pix$.}
\par
The lack of total power observations leads to flux losses, due to the filtering of the large-scale emission which is particularly important for extended sources such as the core we simulate. Focussing on run 4, we estimate that between $\approx 20$ and 65\% of the flux in a $15''$ area around the core is recovered, depending on the tracer. This is in line with simulations regarding the filter-out effect. For instance, \cite{Plunkett23} found that up to 90\% of the original flux can be lost in extended sources when single-dish data are not available.

\begin{table}[!h]
    \centering
    \renewcommand{\arraystretch}{1.4}
    \caption{Parameters used in \textsc{simobserve}. }\label{tab:ALMA-like}
    \begin{tabular}{cccc}
    \hline
     Line & 12m config.& 12m time & 7m time\\
     \hline     
        \hline
     Cont. & C43-1 & 1.1$\,$h & 8.1$\,$h \\
     \dcop (3-2) & C43-2 &  1$\,$h    &  4.7$\,$h          \\
     \htcop (3-2) & C43-2 &  31$\,$min  &    2.4$\,$h          \\
     \cdo (2-1) & C43-2 &  1$\,$h     &   4.4$\,$h      \\
     \ohhdp     &  C43-1   &  1.8 $\,$h    &  12.9$\,$h    \\
    \hline

\hline
  \end{tabular}
\end{table}

\subsection{Column density computation\label{sec:col_density}}
The different approaches for computing the cosmic-ray ionisation rate depend on the column densities of the involved species. To estimate them, we use the approach of \cite{Mangum15}:
\begin{equation}
N_\mathrm{col} =  \frac{8 \pi Q_\mathrm{rot} (T_\mathrm{ex}) \nu^3}{c^3 A_\mathrm{ul} g_\mathrm{u} } \times \frac{\exp \left ( E_\mathrm{u}/k_\mathrm{B}T_\mathrm{ex} \right)}{ \exp \left ( h \nu/k_\mathrm{B}T_\mathrm{ex} \right) -1} \times \int \tau_\nu {\rm d}V \; ,
\label{Ncol_tau}
\end{equation}

\noindent where $h$, $k_\mathrm{B}$, and $c$ are the Planck constant, the Boltzmann constant, and the speed of light in vacuum; $E_\mathrm{u}$ is the upper-level energy, $g_\mathrm{u}$ the upper-level multiplicity, $\nu$ the line frequency, $A_\mathrm{ul}$ the Einstein coefficient for spontaneous emission, and $ Q_\mathrm{rot} (T_\mathrm{ex})$ the partition function at the excitation temperature $T_\mathrm{ex}$. The values used for the spectroscopic constant are listed in Table \ref{tab:transitions}, and they are taken from the CDMS catalog\footnote{ \url{https://cdms.astro.uni-koeln.de/classic/}.}. The partition functions are from  \cite{Giannetti19} for \ohhdp, the CDMS catalogue for \cdo and \htcop, and \cite{Redaelli19} for \dcop. To compute the partition function at the requested temperature, the values have been linearly interpolated, when necessary. $\int \tau_\nu {\rm d}V$ is the integral along the velocity axis of the optical depth $\tau_\nu$ computed channel by channel using (cf. \citealt{Caselli02b})
\begin{equation}
\tau_\nu = - \ln \left [1 - \frac{T_\mathrm{MB}}{J_\nu (T_\mathrm{ex}) - J_\nu (T_\mathrm{bg})} \right ]    \; ,
\label{eq:tau}
\end{equation}
where $T_\mathrm{MB}$ is the line main beam temperature, $J_\nu (T)$ is the equivalent Rayleigh-Jeans temperature at the line frequency\footnote{We highlight that the computation should be performed using the frequency of each channel. However, we only focus on small frequency/velocity coverage ($2\,$\kms), and therefore the error introduced by using the transition frequency is of the order of $10^{-6}$, negligible for our results.}, and $T_\mathrm{bg}=2.73\, \rm K$ is the background temperature. The obtained optical depth profiles are integrated along the velocity range $[-1:1]\,$\kms, which is large enough to include the line emission in all the synthetic cubes analysed for this work. \par
The excitation temperature for all the transitions is $T_\mathrm{ex} = T_\mathrm{k} =15\, \rm K$ when assuming LTE conditions. For the two non-LTE cases, the excitation temperatures have been selected based on the critical densities of the analysed transitions and on available literature data.\footnote{\textsc{polaris} does not automatically return the excitation temperature, which in any case is a quantity defined in each Voronoi cell. Computing average values from its 3D distribution is not straightforward (as discussed in \citealt{Redaelli19}).}. The \cdo first two rotational lines have relatively low critical densities ($n_\mathrm{c}\approx 10^3 \, \rm cm^{-3}$), and it is hence reasonable to assume that they are thermalised by collisions with H$_2$, leading to $T_\mathrm{ex} = T_\mathrm{k} = 15\, \rm K$. The critical density of \ohhdp is higher ($n_\mathrm{c}\approx 10^5 \, \rm cm^{-3}$, \citealt{Hugo09}), and the line is likely sub-thermally excited {, leading to $T_\mathrm{ex}<T_\mathrm{k}$. We adopt $T_\mathrm{ex} = 10\, \rm K$, which is frequently employed in the literature. For instance, \citet{Caselli08} computed $T_\mathrm{ex} = 7-13 \,\rm K$ in the envelope of protostellar cores that have gas temperatures of $10-15\, \rm K$, close to that of our simulations; \cite{Friesen14} adopted $T_\mathrm{ex} = 12\, $K; \cite{Redaelli21a, Redaelli22} used $T_\mathrm{ex} = 10\,$K. \dcop and \htcop are isotopologues with similar critical densities, and it is reasonable to assume that the same rotational transitions share similar excitation temperatures. However, literature information about these are scarce. Using a full non-LTE modelling of the \dcop lines in the well-known core L1544, \cite{Redaelli19} found $T_\mathrm{ex}(1-0) = 5.7\, \rm K$ and $T_\mathrm{ex}(3-2) = 7.8\, \rm K$. L1544 is, however, colder than our simulated cores. We have hence used the online tool RADEX\footnote{available at \url{http://var.sron.nl/radex/radex.php} \citep{vandertak07}.} to confirm these values. Using $n=10^5\, \rm cm^{-3}$, $T_\mathrm{k} = 15\, \rm K$, and $N_\mathrm{col} = 10^{12}\, \rm cm^{-2}$, we derived $T_\mathrm{ex} \approx 8-12\, $K for the lower-J transitions and  $\approx 5\,$K for the higher-J ones. We therefore set $T_\mathrm{ex}(1-0) = 10\, \rm K$ and $T_\mathrm{ex}(3-2) = 5.5\, \rm K$ for both isotopologues. In Appendix~\ref{app:tex} we show that a 20\% variation of these values does not affect our conclusions.} The last column of Table~\ref{tab:transitions} summarises the excitation temperature values used in the LVG analysis. 
\par
{  To estimate the uncertainties on the derived column density values ($rms_N$), we apply standard error propagation on Eq.~\eqref{Ncol_tau}, assuming that the frequency channels are independent and using the small-error approximation. Then, the uncertainty propagation leads to
\begin{equation}
\begin{aligned}
{rms_N =} & {  \frac{8 \pi Q_\mathrm{rot} (T_\mathrm{ex}) \nu^3}{c^3 A_\mathrm{ul} g_\mathrm{u} } \times \frac{\exp \left ( E_\mathrm{u}/k_\mathrm{B}T_\mathrm{ex} \right)}{ \exp \left ( h \nu/k_\mathrm{B}T_\mathrm{ex} \right) -1} \times rms } \\ 
&{\times \Delta V_\mathrm{ch} \sqrt{\sum_{k=ch_{i}}^{ch_{ f}} \left ( \frac{1}{J_\nu (T_\mathrm{ex}) - J_\nu (T_\mathrm{bg}) - T_\mathrm{MB}^k}   \right )^2}\;} ,
\end{aligned}
\label{eq:errNcol}    
\end{equation}

%

where $ch_{ i}$ and $ch_{ f}$ are channels corresponding to the velocity interval over which Eq.~\eqref{eq:tau} is computed, $T_\mathrm{MB}^k$ is the intensity of the $k$-th channel, and $\Delta V_\mathrm{ch}$ is the channel width (in \kms).
}
\par
To estimate the abundances, we derive the total gas column density map from the continuum map as
\begin{equation}
N (\mathrm{H_2}) =     f \times \frac{S_\mathrm{pix} D^2 } {B_\nu(T_\mathrm{dust})\kappa_\nu \, \mu_{\rm H_2} \,  m_\mathrm{H} \Omega_\mathrm{pix}} \; ,
\label{eq:NH2}
\end{equation}
where $f=100$ is the gas-to-dust mass ratio \citep{Hildebrand83}, $D$ is the source distance, $B_\nu(T_\mathrm{dust})$ is the Planck function at the dust temperature $T_\mathrm{dust} = 15\, \rm K$, $\mu_{\rm H_2} = 2.8$ is the mean molecular weight per hydrogen molecule, $m_\mathrm{H}$ is the mass of the hydrogen atom, $S_\mathrm{pix}$ is the flux (in units of $\rm Jy \, pix^{-1}$),  $\Omega_\mathrm{pix}$ is the pixel size (in physical units), and $\kappa_\nu$ is the dust opacity. For the latter, we use the output of \textsc{polaris}, which tabulates the opacities at the simulated wavelength: $\kappa_{\rm 345 GHz} = 0.388 \, \rm cm^2 \, g^{-1} $. Uncertainties on the total column densities are estimated using Eq.~\eqref{eq:NH2}, with the flux noise level of the continuum map as $S_\mathrm{pix}$.

\begin{figure*}[!h]
    \centering
    \includegraphics[width=.9\textwidth]{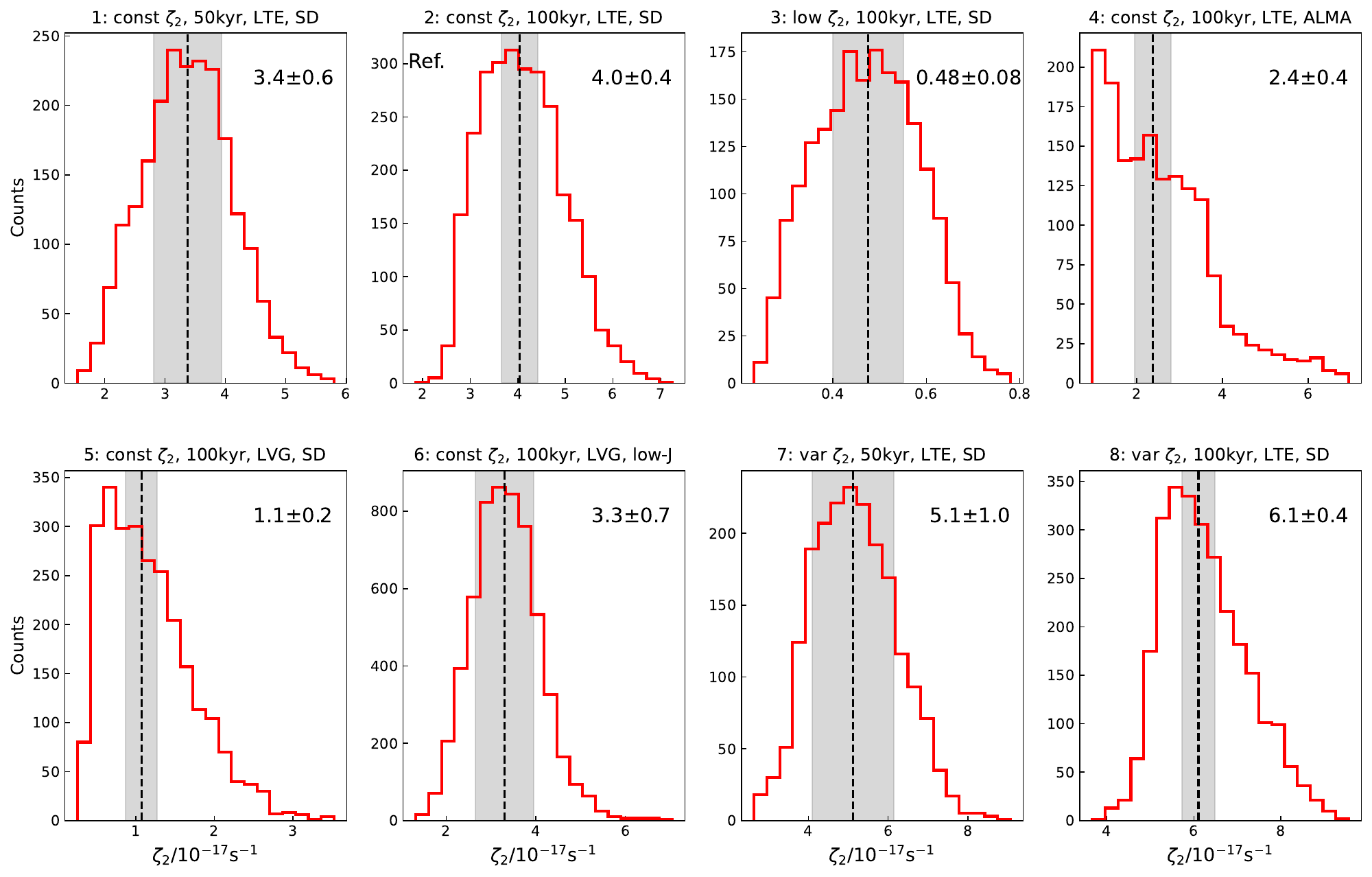}
    \caption{Histogram of the distribution of \crir (in units of $10^{-17}\, \rm s^{-1}$) of the runs as labelled at the top of each panel. We represent the core region where $N({\rm H_2})> 0.5 \times  \text{max}(N({\rm H_2}))$ (i.e. the lowest contour in the panels of Figs.~\ref{fig:tests1_4}, \ref{fig:crir_test5_6}, and \ref{fig:crir_var_sim}). The median value, together with the median uncertainty, is reported in the top-right corner of each panel and shown as a vertical dashed line with a grey-shaded area. Cf. Table~\ref{tab:tests}.}
    \label{fig:crir_histograms}
\end{figure*}

\section{Resulting \crir maps \label{sec:results}}
We now apply the analytical expressions described in Sect.~\ref{sec:ana_approaches} to infer the \crir maps. {Uncertainties on derived quantities are computed pixel-per-pixel assuming standard error propagation calculated from the uncertainties on the column densities that are considered independent}. We neglect, for instance, any source of uncertainty from the reaction rates. { Once the errors are computed, we mask pixels where the signal-to-noise ratio is $\rm S/N \leq 3$.}

\subsection{The \cite{Bovino20} method\label{sec:results_bovino}}
To test the analytical method of \citet[][BFL20]{Bovino20}, we computed \crir in the eight runs described in Table~\ref{tab:tests}, varying the \crir model (constant or variable), the type of radiative transfer (LTE or LVG), the post-processing method (single-dish or ALMA-like), and the frequency of the molecular lines. All the simulations are isothermal at $15\, \rm K$, and, therefore, the rate coefficient in Eq.~\eqref{eq:crir_observables} is $ k_\mathrm{CO}^\mathrm{oH_3^+}  = 2.3 \times 10^{-9}\, \rm cm^{3} \, s^{-1}$.
 \par
In the following subsections, we discuss in detail the results of each run, showing the obtained \crir maps. In order to compare these values with the actual ones, we make use of the \crir distributions in Fig.~\ref{fig:crir_histograms}. Its panels show the histograms of the ionisation rate values extracted in the densest region of the core (i.e. where the H$_2$ column density is higher than 50\% of its peak value), where the signal-to-noise ratio is higher. The medians of the distributions (vertical dashed lines) are directly compared with the actual value of $2.5 \times 10^{-17} \, \rm s^{-1}$ in the runs with constant \crirs. In runs with variable \crirs, we compare pixel-by-pixel the ratio between actual and retrieved values, as discussed in Sect.~\ref{sec:discussion}.
\begin{figure*}[!h]
    \centering
    \includegraphics[width=.75\textwidth]{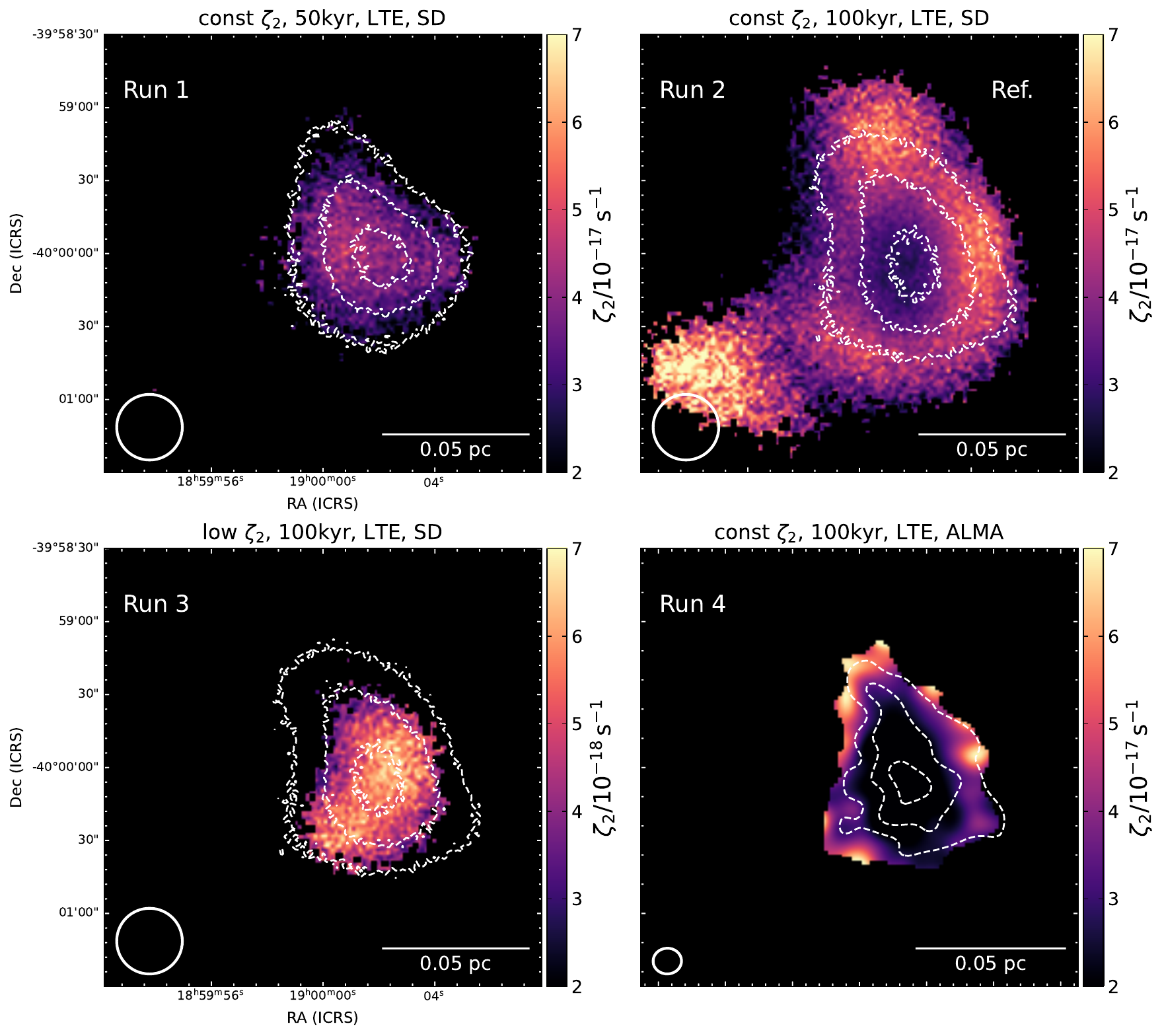}
    \caption{Resulting \crir maps (in units of $10^{-17} \, \rm s^{-1}$) obtained with Eq.~\eqref{eq:crir_observables} in runs 1 to 4. The run ID is reported in the top-left corner, and the key parameters are included at the top of each panel. Run~2 is taken as a reference throughout the rest of this work. The white contours show the 50, 70, and 90\% of the $N\rm (H_2)$ peaks, which are: $6.67\times 10^{22} \,\rm cm^{-2}$ (core in single-dish like analysis at 100$\,$kyr); $6.86 \times 10^{22} \, \rm cm^{-2}$ (core in single-dish-like analysis at 50$\,$kyr); $2.51\times 10^{22} \, \rm cm^{-2}$ (core in ALMA-like analysis at 100$\,$kyr). The beam size and scalebar are shown in the bottom corners of each panel. Note that we show a zoom-in of the central $0.15\,\rm pc$. }
    \label{fig:tests1_4}
\end{figure*}

\subsubsection{Runs 1-2: constant \crirs and single-dish like analysis \label{sec:run1_2}}
The first runs we test have constant $\zeta_2^\mathrm{s} = 2.5 \times 10^{-17} \, \rm s^{-1}$, coupled with LTE radiative transfer of the high-J transitions and single-dish-like analysis. We explored two evolutionary stages, 50$\,$kyr (run 1) and 100$\,$kyr (run 2, the reference run). The resulting \crir maps are shown in Fig.~\ref{fig:tests1_4}, top row. We employ Eq.~\eqref{eq:crir_observables} with $L=0.3$\,pc, which represents the path length (along the line of sight) over which the column densities are integrated. In our case, this corresponds to the size of the simulated box ($\rm 0.3 \,pc \times 0.3 \,pc \times 0.3 \,pc $). A detailed discussion on how to choose $L$ and the derived uncertainty is presented in Sect.~\ref{sec:realObs} and \ref{sec:discussion}. 
\par
The obtained \crir values span the range $\approx (2-8)\times 10^{-17} \, \rm s^{-1} $, with medians in the denser parts of the core of $\zeta_2 = {(3.4 \pm 0.6)}\times 10^{-17} \, \rm s^{-1} $ (50$\,$kyr) and $\zeta_2 = (4.0 \pm 0.4)\times 10^{-17} \, \rm s^{-1} $ (100$\,$kyr), as shown in Fig.~\ref{fig:crir_histograms}. These values should be compared with the actual $\zeta_2^\mathrm{s} = 2.5 \times 10^{-17} \, \rm s^{-1}$. We conclude that, in these runs, the BFL20 reproduces the \crirs within a factor $1.5-1.6$ on average. \par
Concerning the morphology of the retrieved \crir maps, the one at $50 \, \rm kyr$ shows a smaller scatter around the median value than the run at 100$\,$kyr (see Fig.~\ref{fig:crir_histograms}), mainly because we can infer \crir only for positions where $N\rm (H_2) > 4.8 \times 10^{22} \, \rm cm^{-2}$. This is because at this early stage, the \ohhdp abundance is at most $X(\text{\ohhdp}) = 7 \times 10^{-11}$, producing a line peak intensity of $0.5\, \rm \, K$\footnote{The line weakness is the reason why, for this transition, we inject a noise with $rms=50 \, \rm mK$ in the datacube.}. For comparison, at 100$\,$kyr, the \ohhdp abundance reaches $4 \times 10^{-10}$, and the transition is as bright as $3\, \rm K$. This limits the area where \crir is computed with $\rm S/N >3$ in run 1. 
\par
The histogram from run 2 spans a larger range of values than run 1 and presents a tail at higher values, because the retrieved \crir map presents an increase with increasing distance from the core centre, especially in the northern and western directions (cf. top-right panel of Fig.~\ref{fig:tests1_4}). A further enhancement up to $\zeta_2 \sim 9 \times 10^{-17} \, \rm s^{-1}$ is visible in the south-eastern part of the source (note that this does not affect the histogram, which focuses on the high H$_2$ column density region to improve the $\rm S/N$). In Sect.~\ref{sec:discussion} we discuss more in detail the implication of the spatial trends seen in the \crir maps.

\subsubsection{ Run 3: low \crirs and single-dish like analysis \label{sec:run3}}
{  To further expand the range of input \crirs values explored and to be consistent with the tests performed by \cite{Bovino20}, we analysed an additional simulation where the \crir is kept constant on the value $2.5\times 10^{-18} \, \rm s^{-1}$ (low \crirs case). We consider the evolutionary time of $100\, \rm kyr$. The radiative transfer is performed as described in Sect.~\ref{sec:polaris}, adopting LTE conditions and focusing on the high-J transitions. The post-processing is single-dish-like, with a convolution beam size of $27''$. The setup, hence, is identical to the reference run n.~2 except for the input \crirs value and the injected noise level. With this \crirs value, the deuteration process is slow, and the abundances of deuterated species (\hhdp, \dcop) at $100\, \rm kyr$ are orders of magnitudes lower than in the tests with $\text{\crirs} = 10^{-17} -10^{-16} \rm s^{-1}$. We reduce the simulated noise level to $rms=1\,\rm mK$ in the post-processing, to compute the column density of all species significantly.\par
Using the BFL20 method, we obtain the map shown in the bottom-left panel of Fig.~\ref{fig:tests1_4}. The corresponding histogram of the distribution of values is shown in Fig.~\ref{fig:crir_histograms}. The computed values are in the range $(2-8) \times 10^{-18} \rm \, s^{-1}$, and the median value of $(4.8 \pm 0.8 )\times 10^{-18} \rm \, s^{-1}$ is less than a factor of two higher than the input \crirs. }

\subsubsection{Run 4: ALMA-like analysis}
We now focus on the case with constant $\zeta_2^\mathrm{s} = 2.5 \times 10^{-17} \rm \, s^{-1}$ and an ALMA-like {post-processing} as described in Sect.~\ref{alma_analysis}. It is important to discuss the chosen value of $L$. In the single-dish-like analysis, the integrated intensity (or optical depth) is computed along the whole simulated line of sight, i.e. along the whole length of the simulation box (0.3$\,$pc). In the ALMA-like analysis, on the other hand, this is not the case. Once the \textsc{simobserve} task is run, the interferometer acts as a low spatial-frequency filter, { and the emission over scales larger than the so-called maximum recoverable scale ($\theta_\mathrm{mrs}$) is filtered out. We hence select the $\theta_\mathrm{mrs} = 15''$ that the ALMA OT predicts for the chosen antenna configuration in the \olineh setup. This corresponds to $L=0.15\, \rm pc$.} \par
The resulting \crir maps {is shown in the bottom-right panel of Fig.~\ref{fig:tests1_4}.} 
The histogram of {run 4} (top-right panel of Fig.~\ref{fig:crir_histograms}) is asymmetric, with a global maximum at low \crir values ($1 \times 10^{-17} \, \rm s^{-1}$), and a tail up to $7 \times 10^{-17} \, \rm s^{-1}$. This is due to the spatial gradient seen in the bottom-right panel of Fig.~\ref{fig:tests1_4}, where \crir increases as $N\rm (H_2)$ decreases. In the central part of the core, the actual \crirs is {well recovered}, as confirmed by comparing the median {value ${ \langle \zeta_2 \rangle = (2.4 \pm 0.4)\times 10^{-17} \, \rm s^{-1}}$ with $\zeta_2^\mathrm{s} = 2.5 \times 10^{-17} \, \rm s^{-1}$.} 
\begin{figure*}[!h]
    \centering
    \includegraphics[width=\textwidth]{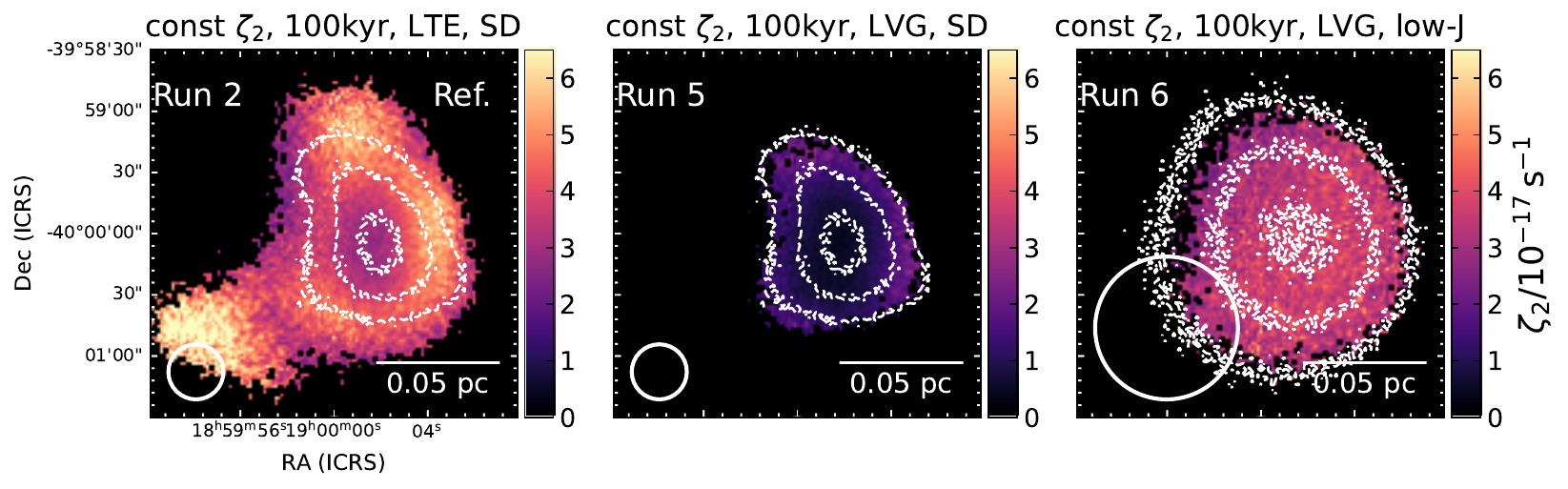}
    \caption{Summary of runs performed in LVG approximation, compared with the reference case (run 2, left panel). The central panel refers to the analysis performed on the high-J transitions for \dcop, \htcop, and \cdo, whilst the low-J ones are used in the map in the right panel. The white contours are the same as in Fig~\ref{fig:tests1_4}. Note that the colour scale is the same across all the panels. The beam size and scalebar are shown in the bottom left and right corners of each panel.}
    \label{fig:crir_test5_6}
\end{figure*}

\begin{figure*}[!h]
    \centering
    \includegraphics[width=.7\textwidth]{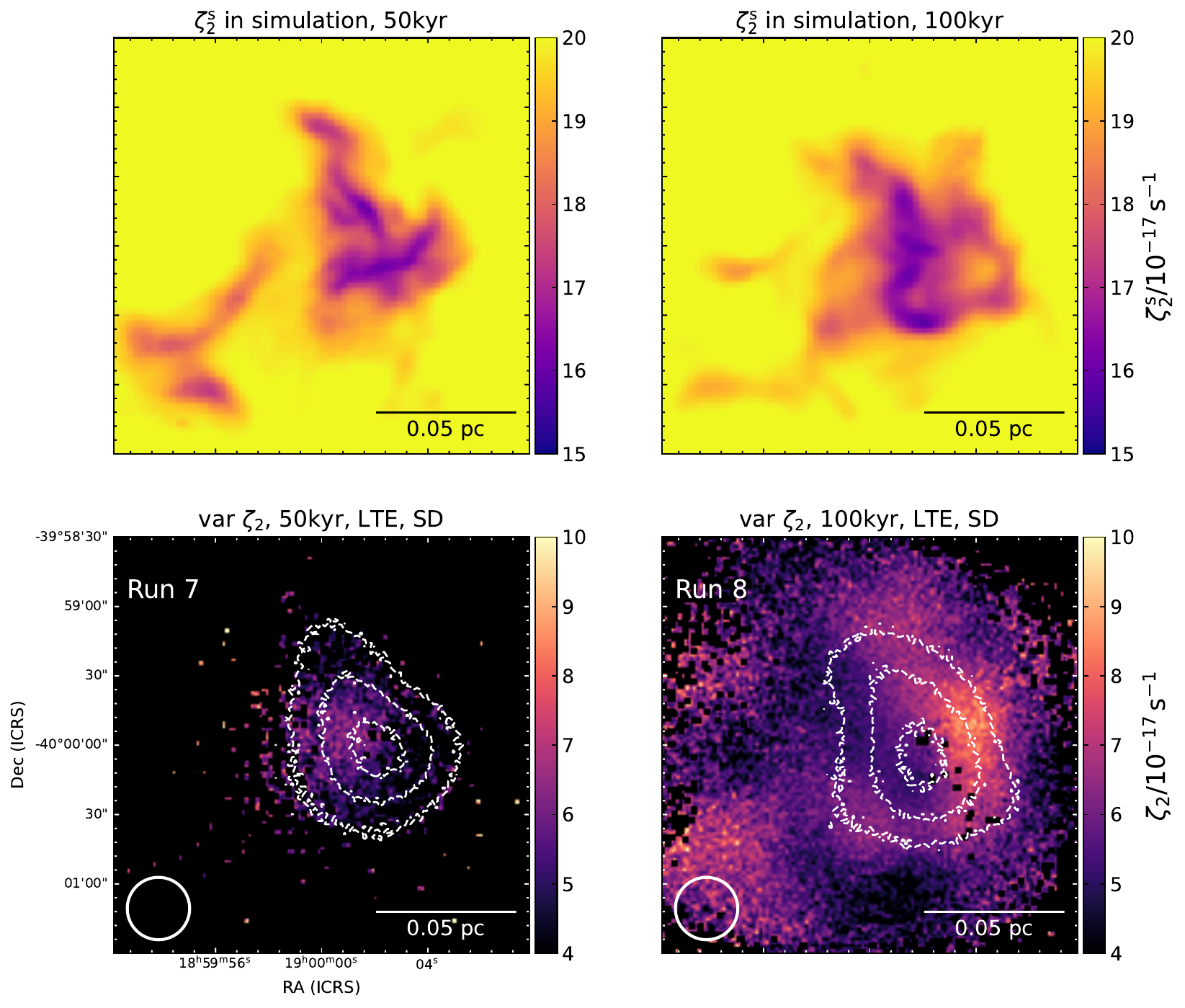}
    \caption{Top row: maps of the density-averaged \crirs in the simulation with variable cosmic rays, at 50$\,$kyr (left) and 100$\,$kyr (right). Bottom row: same as Fig.~\ref{fig:tests1_4}, but for runs 7 and 8. Note the colour scale is the same within rows, but it changes from the top panels to the bottom ones.}
    \label{fig:crir_var_sim}
\end{figure*}
\subsubsection{Runs 5-6: LVG radiative transfer and low-J transition \label{sec:run5_6}}
Runs~5 and 6 explore the effects of the type of radiative transfer performed and of the rotational levels of the lines used in the analysis. Both runs use the LVG option in \textsc{polaris}. Run~5 employs the \dcop, \htcop, and \dcop transitions at frequencies $\approx 215-260 \, \rm GHz$. The line intensities are generally lower than in the corresponding LTE calculation. The change is the smallest for the \cdo (2-1) line ($\approx 15\%$), which is expected as this transition is thermalised. On the contrary, the \dcop and \htcop (3-2) fluxes are reduced by a factor of up to 3 and 10, respectively. In fact, due to their high critical densities, these transitions are subthermally excited. This requires reducing the simulated noise in this run to 50$\,$mK, to improve the final S/N. Table~\ref{tab:transitions} reports the excitation temperatures used to compute the column densities, following what {is} stated in Sect.~\ref{sec:col_density}.
\par

The resulting \crir map is shown in the central panel of Fig.~\ref{fig:crir_test5_6}, and it spans values in the range $(0.5-2.5 )\times 10^{-17} \, \rm s^{-1}$. The median in the high-density part of the core (see Fig.~\ref{fig:crir_histograms}) is $\langle \zeta_2 \rangle = (1.1 \pm 0.2)\times 10^{-17} \, \rm s^{-1}$, hence a factor $2.3$ smaller than the actual \crirs value of the simulations. The \crir distribution is asymmetric, as a consequence of the increasing trend of \crir with decreasing $N \rm (H_2)$ that has also been noted in the previous runs at 100$\,$kyr.
\par

In run~6, we explore the effect of targeting the low-J transitions of \cdo, \htcop, and \dcop,  {as these lines are often targeted by spectroscopic surveys at $3\,$mm. We adopt a large beam size of $\approx 70''$, to simulate unresolved observations}. The resulting \crir map, shown in Fig.~\ref{fig:crir_test5_6}, presents the flattest distributions of values, with the smallest scatter around the median (see Fig.~\ref{fig:crir_histograms}). This happens because the area where we recover the \crir map is comparable to the beam size, and hence any spatial trend is smoothed out by the poor resolution. The resulting median $\langle  \zeta_2 \rangle  = (3.3 \pm {0.7}) \times 10^{-17} \, \rm s^{-1} $ agree with the actual \crirs $= 2.5 \times 10^{-17} \, \rm s^{-1}$ within a factor of $ 1.3$.

\subsubsection{Runs 7-8: variable \crirs \label{Sect:var_crir}}
Runs 7 and 8 employ a variable \crirs, as described in Sect.~\ref{simulations}. In this case, it is not straightforward to compare the resulting \crir maps with the simulation value, which is a three-dimensional, spatially-dependent quantity. For this comparison, we  {compute} the line-of-sight density-averaged \crirs at timesteps 50 and 100$\,$kyr. The results are shown in the top row of Fig.~\ref{fig:crir_var_sim}. The maps show that the cosmic-ray ionisation rate decreases from $\approx 2 \times 10^{-16} \, \rm s^{-1}$ at low densities down to  $\approx 1.5 \times 10^{-16} \, \rm s^{-1}$ in the core's centre, with a variation of 25\%. Moreover, the average value in these simulations is almost one order of magnitude larger than in those with constant \crir, offering us the chance to test a high \crir case. \par

The maps of \crir computed using Eq.~\eqref{eq:crir_observables} are shown in Fig.~\ref{fig:crir_var_sim} (bottom panels). In general, they tend to underestimate the simulation values, and the disagreement is larger at the earlier timestep. Overall, our results underestimate the actual \crirs of a factor $\approx 3$. Concerning the decrease of \crir with increasing total column density, we note that at 50$\,$kyr the extension of the retrieved \crir map corresponds to only a few beams, and no clear spatial trend is seen. In the later timestep, a positive gradient is visible from the core's centre to the westernmost outskirts, but no symmetric trend is visible in the other directions. A localised enhancement ($\zeta_2 = 7-9 \times 10^{-17} \, \rm s^{-1}$) is seen in the south-east corner of the source, with no clear counterpart in the actual \crirs map, where instead a local decrease is visible in this area. We conclude that the resulting \crir map does not reproduce the morphology of the actual one, as we further discuss in Sect.~\ref{sec:discussion_morphology}, but provides an accurate average estimate of \crirs. In addition, we note that the \crirs gradient in the original simulations is smaller than the intrinsic error of the analytical formula.

\begin{figure*}[!h]
    \centering
    \includegraphics[width=.75\textwidth]{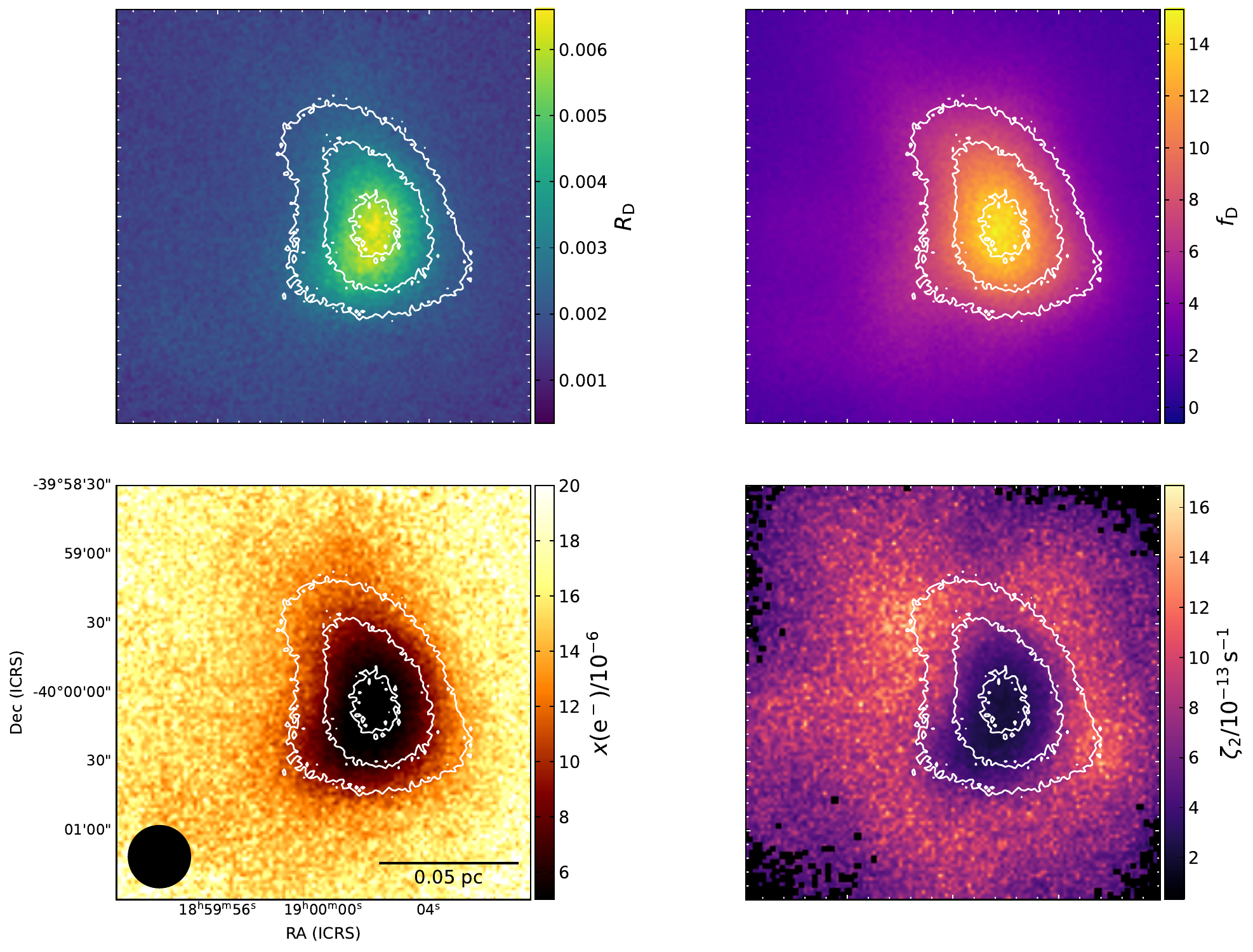}
    \caption{Maps of the key quantities employed by CWT98: \dcop deuteration level (top-left panel), CO depletion factor (top-right), electronic fraction (bottom-left), and \crir (bottom-right).  Note that the colour scales for the last two quantities are in units of $10^{-6}$ and $10^{-13} \, \rm s^{-1}$, respectively. These maps assume the molecular column density and total gas column density derived in run~2. {The mean uncertainties are $2.6 \times 10^{-4}$ ($R_\mathrm{D}$), $0.3$ ($f_\mathrm{D}$), $ 3\times 10^{-6}$ ($x_\mathrm{e^-}$), and $1.5\times 10^{-13}\, \rm s^{-1}$ (\crir).} The beam size and scalebar are shown in the bottom-left panel. The contours show $N \rm (H_2)$,  as in Fig.~\ref{fig:tests1_4}.
    \label{fig:crir_test2_caselli}}
\end{figure*}

\subsection{The \cite{Caselli98} analytical method\label{sec:analysis_caselli98}}
We now focus on the analytical method proposed by \citet[][CWT98 hereon]{Caselli98}  to test the limitations already discussed in \cite{Caselli02b}, providing robust evidence via an accurate methodology. The CWT98 approach has the advantage of depending on commonly observed tracers. Note that the equations depend on the H$_2$ volume density, which we estimate as $n(\mathrm{H_2}) = N(\mathrm{H_2})/L$, where $L$ is set on the same value used for the BFL20 method for each run. 

\subsubsection{The reference run 2}
For the sake of observational applicability, we have tested the behaviour of the original equations (Eq. 3 and 4 of CWT98). 
 {Here}, we present the results for our reference case (run 2). In Fig.~\ref{fig:crir_test2_caselli}, we show the relevant required quantities, in particular, the deuteration level of \hcop (top left panel), and the CO depletion factor
\begin{equation*}
   f_\mathrm{D} = \frac{X^{\rm st}(\mathrm{CO})}{X(\mathrm{CO})} \; ,
\end{equation*}
where $X^{\rm st}(\mathrm{CO}) = 1.2 \times 10^{-4}$ is the CO standard abundance. The \hcop deuteration level peaks at $R_\mathrm{D} = {(6.9 \pm 0.3)}\times 10^{-3}$ towards the core's centre, where the CO depletion reaches $f_\mathrm{D} = {( 15.8 \pm  0.6 )}$. Hence, the deuteration level is smaller than the values spanned by the cores of CWT98, but it fulfils the requirement $R_\mathrm{D} < 0.023 f_\mathrm{D}$ under which the equations can be applied. We further discuss this in Sect.~\ref{sec:discussion}.  The derived values for the electronic fraction, shown in the bottom-left panel, are in the range $(1-10) \times 10^{-6}$. These are overestimated by more than two orders of magnitude compared to the original simulations, where $x(\mathrm{e^-}) \approx \text{a few} \times 10^{-8}$. This error propagates to the resulting \crir map (bottom-right panel of Fig.~\ref{fig:crir_test2_caselli}). The equation overestimates the original \crirs value by more than three orders of magnitudes, especially at the core's outskirts. \par

The original method made several simplifications and assumptions, such as, for example, the reaction rates at constant temperature (10$\,$K), and the lack of ortho- and para-state separation. Furthermore, several reaction rates have been updated since then. We have, therefore, derived the equations again using the same theoretical approach of CWT98, but with the formalism of BFL20, to show that the large overestimates produced by the method are not ascribed to these parameters but rather to the approximations made to obtain the formula. The derivation is illustrated in Appendix \ref{app:Paola}. We have then computed the electronic fraction and the cosmic-ray ionisation rate using the updated set of equations~\eqref{eq:PaolaNew}. We set the HD abundance to $X( \rm HD) = 1.5 \times 10^{-5}$ \citep{Kong15}, and the para-fraction of H$_3^+$ to $f_\mathrm{para} = 0.7$. The latter is consistent with estimates of this parameter in diffuse clouds (see e.g., \citealt{Crabtree12}, and references therein). These values have also been verified in the simulation, and they agree within less than a factor of two (see also \citealt{Lupi21}). \par
\begin{figure}[!h]
    \centering
    \includegraphics[width=0.35\textwidth]{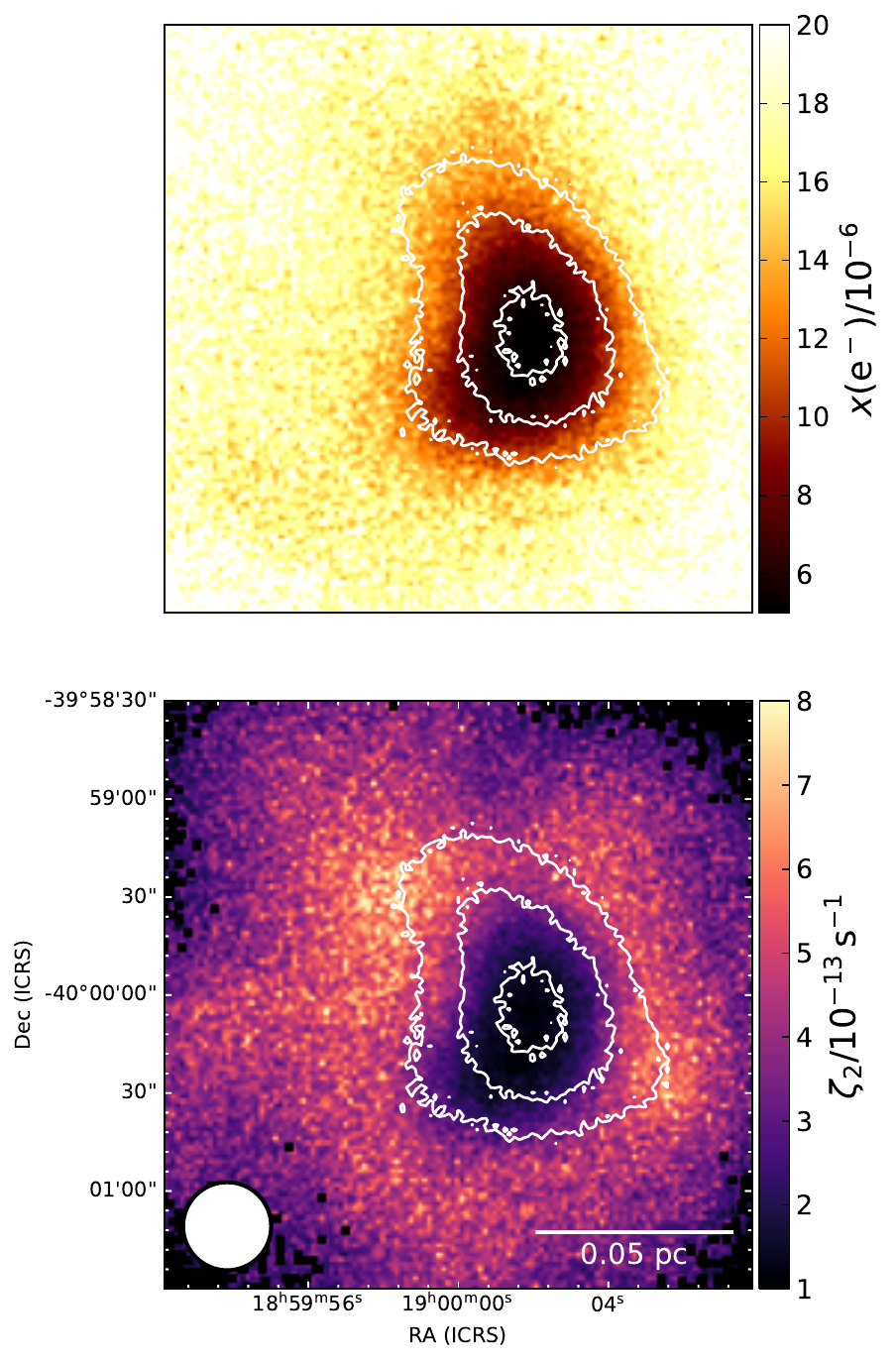}
    \caption{Electronic fraction (top panel) and \crir map (bottom panel) from the reference run~2, using the updated formulation of CWT98. The contours show H$_2$ column density as in Fig.~\ref{fig:tests1_4}. The {mean} uncertainties are $\mathbf {3}\times 10^{-6}$ for $x (\mathrm{e^-})$ and ${0.8} \times 10^{-13} \, \rm s^{-1}$ for \crir. The beam size and scalebar are shown in the bottom corners of the lower panel. }
    \label{fig:crir_casellinew}
\end{figure}

The resulting maps are shown in Fig.~\ref{fig:crir_casellinew}. Towards the core's centre, the $x(\rm e^-)$ values are 15-20\% lower than those derived with the original equations, but still strongly overestimated. As a consequence, in this area, the new estimates for \crir are a factor of $\approx 2$ lower than those from the original derivation, but we still find $\zeta_2 \approx (1.5\pm {0.2})\times 10^{-13} \rm s^{-1}$, i.e. more than three orders of magnitude higher than the actual value.

\subsubsection{CWT98 results in all tested runs}
\begin{figure*}[!h]
    \centering
    \includegraphics[width=.9\textwidth]{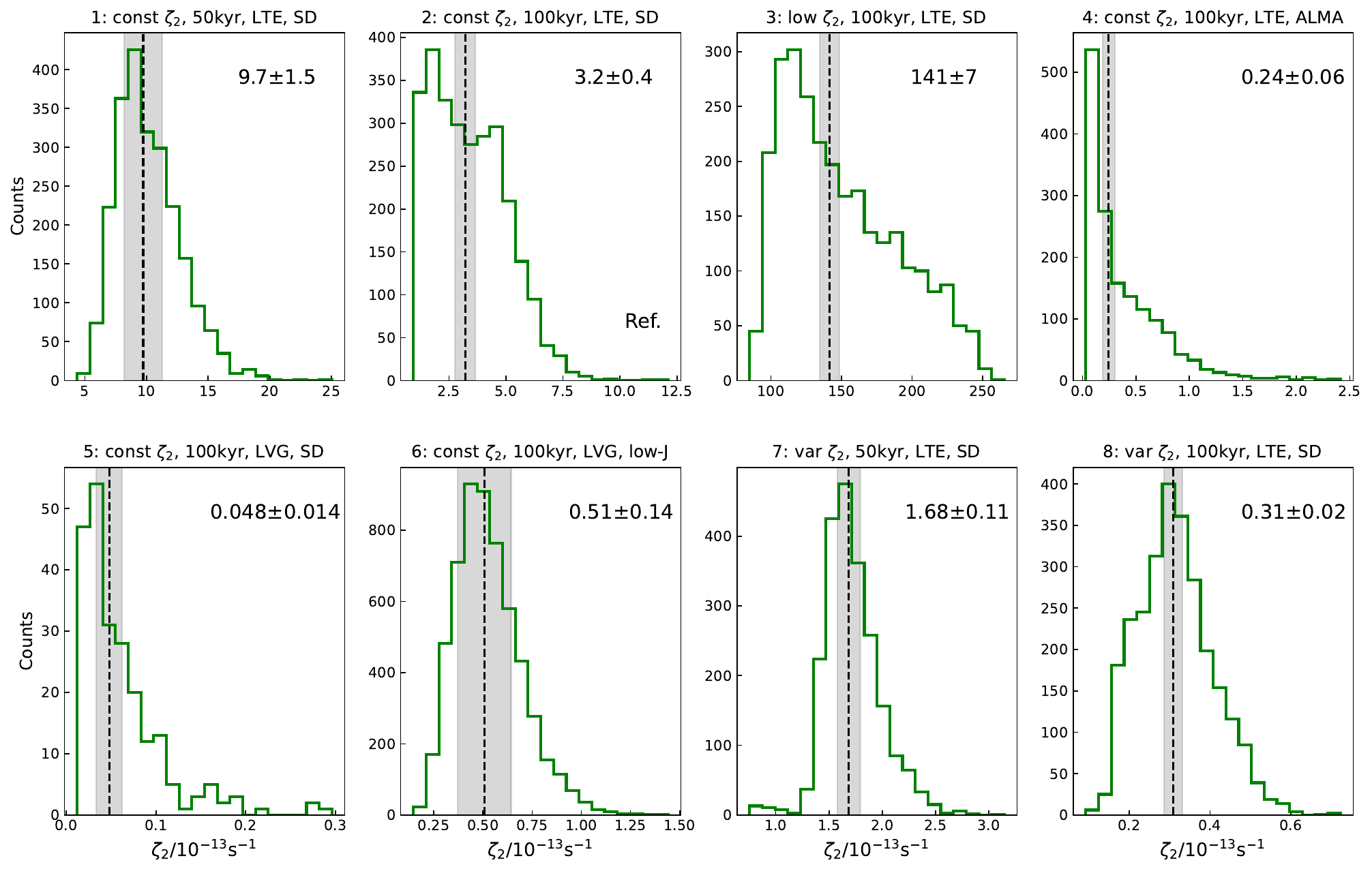}
     \caption{Same as Fig.~\ref{fig:crir_histograms}, but the ionisation rate is estimated using the new formulation of the CWT98 method presented in Appendix~\ref{app:Paola}, summarised in Eq.~\eqref{eq:PaolaNew}. Note that the \crir values are normalised to $10^{-13} \, \rm s^{-1}$ in all panels. The median ($\pm$ median uncertainty) is reported in the top-right corner {and shown with the vertical dashed line and shaded area in each panel.}. The reference run is labelled with ``Ref'' in the bottom-right corner.
    \label{fig:histogram_casellinew}}
\end{figure*}

We now describe the behaviour of the CWT98 method applied to all the remaining runs. We adopt the new formulation of the method, described in Appendix~\ref{app:Paola}. The histograms of the resulting maps, focussing on the central part of the core, are presented in Fig.~\ref{fig:histogram_casellinew}. The most notable feature is that in all tests the retrieved values overestimate the actual ones.{ The median values in the runs performed on the simulation using $\zeta_2^\mathrm{s} = \text{cost} = 2.5 \times 10^{-17} \, \rm s^{-1}$ range from $(4.8\pm 1.4)\times 10^{-15} \, \rm s^{-1}$ (run~5) to  $(9.7\pm 1.5)\times 10^{-13} \, \rm s^{-1}$ (run~1), i.e. an overestimation of two to four orders of magnitude. In the low \crir case (run~3), we obtain $\langle \zeta_2 \rangle = (1.41\pm 0.07)\times 10^{-11} \, \rm s^{-1}$ (overestimated by almost seven orders of magnitude). The two tests performed on simulations with $\zeta_2^\mathrm{s} = \text{var} \approx 10^{-16} \, \rm s^{-1}$ result in $(1.68\pm 0.11 )\times 10^{-13} \, \rm s^{-1}$ (run 7) and $(3.1\pm 0.2)\times 10^{-14} \, \rm s^{-1}$ (run 8), again overestimating the actual \crir by two-three orders of magnitude.}

\subsection{Comparison of the methods on real observations \label{sec:realObs}}
We now compare the two analytical methods on real observations of prestellar cores found in the literature. We found three sources for which all the needed observables are available: L1544, L183, and TMC-1C. The literature values of the required quantities are listed in Table~\ref{tab:real_obs}. For these cores, the H$_2$ volume density is well characterised, and we used this quantity directly in the CWT98 equation ~\eqref{eq:PaolaNew}. For the BFL20 method, a discussion on the parameter $L$ is required. The physical meaning of $L$ is the length of the path on the line-of-sight along which the column densities are computed; in other words, $L$ is the depth of the emitting source. This, in the simulations, is known by construction. In our setup, we cut a $(0.3\, \rm pc)^3$ subcube in the initial larger simulation box (0.6$\,$pc) which is initialised with molecular gas, hence full of e.g. CO. Emitting gas, therefore, is found along the whole box, and the choice of $L$ to be equal to the size of the subcube is well justified. If we were to cut a smaller subcube, $L$ should be adjusted accordingly, since the depth of the emitting gas would also be reduced (see also Appendix~\ref{app:L_study}) for further details). This does not apply to real observations. Real cores are finite, and the length of the emitting gas is limited along the length of the line of sight. $L$ hence has to be computed as the source size along the line of sight, which however is completely unknown. For isolated prestellar cores, we propose to compute $L$ by considering the 20\% isocontour of the $N \rm (H_2)$ peak. Using this prescription, the sizes of the three analysed cores are $0.20-0.35 \,  \rm pc$. The final \crir values are also reported in Table~\ref{tab:real_obs}.\par
\begin{table*}[!h]
\centering
\renewcommand{\arraystretch}{1.4}
\caption{Comparison of \crir values obtained with the CWT98 and BFL20 methods towards three prestellar cores, assuming a gas temperature of $10\,$K. The reaction rates have been computed at the temperature of $10\,$K.}
\label{tab:real_obs}
\begin{tabular}{c|cccccccc}
\hline
       & $R_\mathrm{D}$ & $R_\mathrm{H}$ & $f_\mathrm{D}$ & $n\rm (H_2)$ & $N \rm (oH_2D^+)$ & $L$ & \crir (CWT98) & \crir (BFL20) \\
       &  $\times 10^{-2}$   &$\times 10^{-5}$  &         &$\times 10^5\, \rm cm^{-3}$ &  $\times 10^{13}\, \rm cm^{-2}$ & pc &      $\times 10^{-14}\, \rm s^{-1}$          &    $\times 10^{-17}\, \rm s^{-1}$            \\
       \hline
L1544  & $ 3.5^a  $      &     $5.9^{e}$     &     $14^f$     &  $14^g$  &    $3.2^h$     & 0.20  &     $2.0$    &     $1.0$          \\
L183   &  $ 5.1^b$       &     $2.3^e$    &      $12^g$     &   $10^g$ &    $2.5^h$     &  0.20 &     $0.2$     &     $0.7$      \\
TMC-1C & $ 1.6^{c,d}$    &     $2.8^{d}$     &     $3^{c}$      &  $4.5^d$     &   $0.9^h$   & 0.35   &   $1.2$   &  $1.7$        \\
\hline
\end{tabular}
\tablefoot{References: $^a$ \cite{Redaelli19}; $^b$ \cite{Juvela02}; $c$ \cite{Schnee07}; $^d$ \cite{Fuente19}; $^e$ \cite{Lattanzi20}; $^f$ \cite{Bacmann02}; $^g$ \cite{Crapsi05}; $^h$ \cite{Caselli08}
}
\end{table*}

The \crir values computed with BFL20 are in the range $(0.7-1.7)\times 10^{-17} \, \rm s^{-1}$, whilst with CWT98 we obtain $(0.2-2.0)\times 10^{-14} \, \rm s^{-1}$. Note that there are many uncertainties in the analysis. For instance, the literature values have been computed with data from different telescopes (hence at different resolutions). Furthermore, we assume $T=10\, \rm K$, whilst some of the sources might be colder \citep[cf.][]{Caselli08}. However, the uncertainties on these quantities cannot explain a difference of three orders of magnitude between the two methods. These examples, hence, confirm that CWT98 tends to produce overestimated results compared to BFL20.\par 
    The actual \crir value in these objects is not known, however \cite{Redaelli21b} derived $3.0\times 10^{-17} \, \rm s^{-1}$ in L1544, and \cite{Fuente19} found  $\zeta_2 = (5-18)\times 10^{-17} \, \rm s^{-1}$ in the translucent cloud associated with TMC1. Both papers used extensive modelling of the sources, coupling chemical models with radiative transfer analysis on a large set of molecular tracers. \cite{Pagani09} explored the chemistry and structure of L183, assuming ionisation rates in the range $(0.1-10)\times 10^{-16} \, \rm s^{-1}$ and, even though a definite best-value for this parameter is not given, the authors used $\zeta_2 = 2 \times 10^{-17}\rm \, s^{-1}$ in their most detailed modelling. Furthermore, the most recent models of CR propagation in the dense gas predict \crir values of at most a few $10^{-16} \rm s^{-1}$, unless a local source of CR re-acceleration is present \citep[cf.][]{Padovani16,Padovani18,Padovani22}. It is safe to assume, in summary, that values as high as $\zeta_2 = 10^{-14}\, \rm s^{-1}$ are excluded for these quiescent and dense cores.

\section{Discussion \label{sec:discussion}}
\subsection{Results and limitations of the methods}
\begin{figure}[!h]
    \centering
    \includegraphics[width=0.48\textwidth]{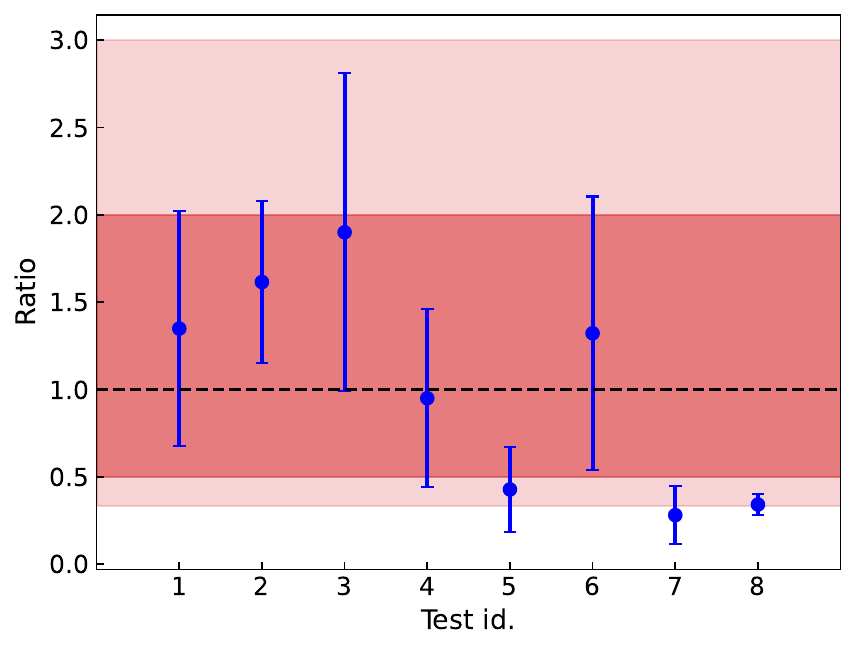}
    \caption{Ratio between \crir retrieved in each run ($x$-axis labels), and the actual value \crirs. The ratios are median computed in each map, in the same regions used to produce the histograms of Fig.~\ref{fig:crir_histograms}. The shaded areas show a variation of a factor of 2 (darker shade) or 3 (lighter shade). The dashed line represents where the ratio is equal to 1. The error bars are 3$\sigma$ median uncertainties. }
    \label{fig:crir_bovino_summary}
\end{figure}

In Fig.~\ref{fig:crir_bovino_summary}, we summarise the results obtained with the BFL20 method in the eight runs of Sect.~\ref{sec:results_bovino}. In the case of constant \crirs, {we show the average ratio between the derived values and the reference $\zeta_2^\mathrm{s} = 2.5 \times 10^{-17} \rm s^{-1}$ (runs~1, 2, and 4 to 6) or $\zeta_2^\mathrm{s} = 2.5 \times 10^{-18} \rm s^{-1}$ (run~3)}. For runs~7 and 8, where the input \crirs is variable, we show the median of the ratio between the derived values and the input values (see top panels of Fig.~\ref{fig:crir_var_sim}). Error bars are computed as three times the median uncertainties over the pixels considered to evaluate the median. As fo the histograms in Fig.~\ref{fig:crir_histograms}, we consider only the densest part of the core. \par
 
Fig.~\ref{fig:crir_bovino_summary} shows that the retrieved values are within a factor of $1.5-3$ from the actual ones. The offset is not constant, nor systematic. In the cases with low and constant \crirs, Eq.~\eqref{eq:crir_observables} overestimate the input value by a factor of up to $1.5$ {(except run~5)}. On the other hand, for the two runs with variable \crirs (runs~7 and 8), the resulting \crir maps tend to underestimate the actual values by a factor of $\approx 3$. {Overall, the BFL20 formula represents a robust and reliable method to estimate the order of magnitude of \crir in dense regions. As for similar analytical methods, even if the BFL20 method shows to be accurate within a factor of 2-3, several aspects should be taken into account when it is applied to actual observations}. The first one is that the analytic expression depends on the column density of four molecular tracers. If any of these is affected by a systematic error, this will propagate to the \crir estimation. Column densities strongly depend on the chosen excitation temperature values. By performing an LTE analysis, we have initially avoided this problem, fixing the $T_\mathrm{ex}$ for all the molecular tracers to the constant gas temperature. In the LVG runs, we selected $T_\mathrm{ex}$ {looking for literature references and checking the selected values with non-LTE tools (such as \textsc{radex}).} Indirectly, this work hence provides good estimates of the $T_\mathrm{ex}$ of several commonly observed transitions, in the considered density ($n \approx 10^5 \rm cm^{-3}$) and gas temperature ($T_\mathrm{K} = 15 \, \rm K$) regimes. In reality, the problem of choosing the correct $T_\mathrm{ex}$ has no straightforward solution, especially in the case of subthermally-excited lines. We strongly suggest, when possible, using multiple lines of the same tracer, which allows us to constrain their excitation conditions better.  \par

{ When using optically-thin isotopologues to infer the total column densities of molecular tracers, particular caution has to be paid to the assumed isotopic ratios, since fractionation processes can lead to significant variation from the elemental isotopic ratios (see, e.g., the discussion of \citealt{Colzi20} on $\rm ^{12}C/^{13}C$). In this work, we avoided this problem by assuming consistent isotopic ratios throughout the postprocessing analysis (cf. Sects.~\ref{simulations} and \ref{sec:polaris}).} 
\par
Another crucial parameter in Eq.~\eqref{eq:crir_observables} is the scale length $L$ employed to obtain the column densities, as discussed in Sect.~\ref{tab:real_obs}. In our analysis, $L$ is known by construction from the size of the simulation box, since this is a subcube cut out of a larger simulation initialised with molecular gas. We justify this choice further in Appendix~\ref{app:L_study}. For observed cores, their size (and, in particular, their depths) are in general unknown. We suggest as a prescription to use the contour where $N\rm (H_2)$ is higher than 20\% of its peak value to estimate $L$. In the three real objects we analysed, this approach leads to \crir values in agreement with estimations of this quantity derived with completely different methods. Note that applying this rule to the reference run~2 results in $L=0.15$\,pc, hence overestimating \crir by a factor $\approx2$, still within the uncertainties of the method. This prescription should not be used on interferometric data with a maximum recoverable scale smaller than the actual source size, as this leads to emission filtering. In these cases, we speculate that $\theta_\mathrm{mrs}$ is a proper scale to estimate $L$. This uncertainty on $L$ also affects the approach of CWT98, where however this is mitigated if the gas volume density is known from other observables.  \par
We highlight how the core fraction where we are able to infer \crir is different in each case (see Figs.~\ref{fig:tests1_4}, \ref{fig:crir_test5_6}, and \ref{fig:crir_var_sim}). This is driven, in particular, by the column density of \ohhdp. At earlier times, the abundance of this species is lower, hence the extension of the core where it is detected is smaller. This highlights the main limitation of this method, i.e. that it relies on the detection of \olineh (see also Sect.~\ref{sec:feasibility} for details).
\par
{On the other hand, the analytic method of CWT98 overestimates \crir by several orders of magnitudes in the cases explored in this work, where it should not be employed (see also the discussion in Sect. 3 of \citealt{Caselli02b}). This is due to the overestimation of the electron fraction caused by the neglected kinetics of \hhdp employed to derive Eq.~\eqref{eq:RD_newrates}, in particular, the reactions producing the doubly- and triply-deuterated forms of H$_3^+$, and other destruction channels involving neutrals. Concerning the importance of $\rm D_2H^+$ and D$_3^+$, \cite{Caselli08} developed analytical equations where the deuteration level of \hcop is expressed in terms of all the deuterated forms of H$_3^+$.} \par

{ Note that \crir, when derived in the reference case with the analytic expression of CWT98 ($10^{-14}-10^{-13} \rm \, s^{-1}$), is one to two orders of magnitude higher than in the original paper (where $\zeta_2 = 10^{-16}-10^{-14} \rm \, s^{-1}$ was found when assuming $f_\mathrm{D} = 5$). This is because the simulated core presents relatively low levels of deuteration and a high depletion factor. In Appendix~\ref{app:Paola}, we show how the CWT98 formula leads to increasingly high \crir values when the deuteration is low, and the CO depletion is high (cf. Fig.~\ref{fig:gelato}). In the reference case (run 2), the core centre presents  $R_\mathrm{D} \approx 0.4-0.7$\% and  $f_\mathrm{D} \approx 10-15$. These are not the typical values observed in CWT98, where most sources present a deuteration fraction of a few per cent \citep[see also][]{Butner95, Williams98}. Note that, given the lack of information available back then on the CO depletion, $f_\mathrm{D}> 5$ was not included in their analysis (catastrophic CO freeze-out was first measured one year later, \citealt{Caselli99}). Furthermore, the assumed temperature $T=15\, \rm K$ is higher than typical gas temperatures observed in low-mass prestellar cores \citep{Bergin06, Crapsi07}. {However, in run~8, with variable \crir at 100$\,$kyr, the core's centre presents $R_\mathrm{D} \approx 2$\% and  $f_\mathrm{D} \approx 3-4$, closer to the properties of the objects analysed by CWT98. The retrieved \crir is $(3.1\pm 0.2)\times 10^{-14} \, \rm s^{-1}$ (see Fig.~\ref{fig:histogram_casellinew}), overestimating the actual \crirs by two orders of magnitudes, confirming the limitation of this approach.}
\par
The relatively low deuteration level of the reference case is due to the assumed initial conditions, in particular the initial H$_2$ OPR. By using constant-density one-zone models, we foundthat when $\text{OPR} = 10^{-3}$ (as reported in the dense and evolved prestellar cores, e.g. \citealt{Kong15}), the \hcop deuteration level increases by about one order of magnitude, and the derived \crir values decrease to $\zeta_2 = 10^{-16}-10^{-15} \rm \, s^{-1}$, still a factor of 10-100 more than the actual value. These tests show that the CWT98 analytic method has a marked dependence on the initial OPR, conversely to BFL20, which is relatively unaffected by this parameter, as already discussed in \cite{Bovino20}. The scope of this work is to retrieve \crir under the typical observational conditions while reproducing the exact physical details of a specific observed object is beyond our aims. The latter was done, for example, by \cite{Bovino21}, where the simulation was designed to reproduce six observed cores in Ophiucus.

\subsection{Morphology of the resulting \crir maps\label{sec:discussion_morphology}}
The obtained \crir maps allow us to comment also on the morphology of the retrieved ionisation rate. By looking at Figs.~\ref{fig:tests1_4} and \ref{fig:crir_test5_6}, it is clear that for most of the runs where \crir is computed in an extended part of the source (i.e. when $N \rm (H_2) \lesssim $50\% of its peak), \crir presents a positive gradient with increasing distance from the core's centre. This is also seen in the histograms in Fig.~\ref{fig:crir_histograms}, where asymmetric tails towards the high \crir values are seen in runs 2, 4, and 5. Since in these runs, the actual \crirs is constant, these spatial trends are not real. These considerations holds also for CWT98 (see Figs.~\ref{fig:crir_test2_caselli} and \ref{fig:crir_casellinew}). The single-dish test performed with the larger beam size (run~6) presents the flattest distribution and the smallest scatter around the median because the core is not spatially resolved. On the contrary, in the tests with variable \crirs, the retrieved maps do not show the gradient with the increasing density present in the simulation (see Sect.~\ref{Sect:var_crir} for more details). We conclude that apparent spatial trends should not be trusted, and averages across the densest regions of the source should be instead considered to express the resulting \crir.

\subsection{Observational feasibility\label{sec:feasibility}}
This work aims to provide observers with reliable methods to estimate \crir in real sources. It is hence important to assess the observability of the proposed tracers. The continuum observations are not particularly challenging and do not represent the limiting aspect of this endeavour. Conversely, the feasibility of the molecular line observations requires a more detailed discussion. \par
We first focus on single-dish facilities. As reported in Table~\ref{tab:transitions}, the \htcop (3-2), \dcop (3-2), ad \cdo (2-1) lines can be observed by APEX, i.e. with the nFLASH230 instrument. Using its observing time calculator\footnote{Version 10.0, available online at \url{http://www.apex-telescope.org/heterodyne/calculator/ns/otf/index.php}.}, and standard input values, we compute that on-source times of $1.0-1.4\, \rm h$ are sufficient to reach $rms =100 \, \rm mK$ in a $3'\times 3'$ FoV and a 0.1$\,$\kms channel. The corresponding (1-0) transitions are covered, for instance, by the EMIR receiver mounted on the IRAM 30m telescope. The time estimator\footnote{Available online at \url{https://oms.iram.fr/tse/}.} predicts that, in winter, $1\, $h of on-source time is enough to reach a sensitivity of $100 \, \rm mK$, with a $3'\times 3'$ FoV and using a 0.1$\,$\kms resolution. 
\par
The \olineh transition is the most challenging in terms of sensitivity, due to its high frequency when compared to the other transitions analysed here. The same requirements in terms of spectral resolution, sky coverage, and sensitivity made for the other lines would lead to $25\, $h of on-source time using the multi-beam LASMA receiver mounted on APEX. However, by relaxing the requirements to a FoV of $2' \times 2'$ (still able to cover the portion of the core where $N \rm (H_2) \gtrsim 2.5 \times 10^{22} \, cm^{-2}$) and downgrading the resolution to $0.15\, $\kms, the on-source time reduces to $7.5\,$h, which is manageable in a small project. The downgraded spectral resolution does not impact the computation of the column density, since the lines are still resolved by at least 3 channels. {The sensitivity required by the low \crir case in run~3 ($1\,$mK), on the other hand, is currently well beyond the capabilities of existing single-dish facilities, even in the case of single-point observations at the centre of the source.} 
\par
The observing times required with ALMA are listed in Table~\ref{tab:ALMA-like}. Considering that several lines can be observed simultaneously, these observations appear feasible. {We conclude that in most cases the observations required to compute \crir are feasable, as proved by the increasing number of campaigns recently reported \citep[cf.][]{Giannetti19,Sabatini20,Redaelli21a,Redaelli22}.}

\section{Summary and conclusions \label{sec:summary}}
In this work, we have tested two analytical methods to retrieve the cosmic-ray ionization rate \crir in dense gas. This has been done using synthetic molecular and continuum data, produced via radiative transfer analysis on a set of three-dimensional simulations that include the chemistry of the involved molecular tracers. {This allows us to evaluate with accuracy the loss of information (and then the accuracy of the method) when simulating realistic observations.} \par

The method of \cite{Caselli98} has several limitations by construction, such as $R_\mathrm{D} < 0.029 \times f_\mathrm{D}$ (to avoid $x(\rm e) < 0$ {in the new formulation derived in} Appendix~\ref{app:Paola}). Furthermore, this analytical approach strongly depends on the H$_2$ initial OPR. This limits its applicability, especially when the OPR is reset to much higher values than those in cold cores by conditions such as shocks or protostellar outflows. In our reference case, this method overestimates \crirs by up to four orders of magnitude. In particular, in tests where the deuteration level is a few \%, hence similar to what is observed in several prestellar cores, the \cite{Caselli98} method overestimates by two orders of magnitude the actual \crirs. 
\par
On the contrary, the method of \cite{Bovino20} is generally accurate within a factor of $2-3$ in retrieving the actual \crirs. Its main limitation is linked to the level of total deuteration, since at late evolutionary stages or at very high densities ($n \gtrsim 10^7 \, \rm cm^{-3}$) \hhdp is converted into doubly and triply deuterated forms, and it is not a reliable tracer of the total $ \rm H_3^+$ abundance anymore. This leads to underestimating the actual \crirs, as already pointed out in the original paper \citep{Bovino20}. 
\par
 As a direct example of the application of the two formulae on observational data, {we explored three well-known prestellar objects, with recent literature data on the quantities involved in the calculation. We showed that the \crir values obtained with BFL20 are in overall good agreement with estimations of the same quantities obtained with non-analytical methods. The results of CWT98 are two to three orders of magnitude higher, as seen also in the tests on the simulations. We highlight, however, that to establish the methodology proposed by \cite{Bovino20}, a statistical sample of observed cores and a proper comparison with theoretical models of CR propagation are needed.}
\par
To conclude, we have discussed the feasibility of the observations necessary to use two commonly employed analytical methods to retrieve \crir. Despite the observational challenges, they are accessible with currently available radio facilities. When the physical structure of a source is well known, coupling a chemical code with radiative transfer using multiple tracers could be employed to infer the cosmic-ray ionisation rate, even if its results might be affected by the parameters' degeneracy. For all the other sources (when this approach is not a viable option), the method of \cite{Bovino20} is a model-independent and reliable analytical method to investigate \crir in dense regions.

\begin{acknowledgements}
The authors acknowledge the referee's comments that led to the manuscript's improvement. ER and PC acknowledge the support of the Max Planck Society. SB is financially supported by ANID Fondecyt Regular (project \#1220033), and the ANID BASAL projects ACE210002 and FB210003. AL acknowledges funding from MIUR under the grant PRIN 2017-MB8AEZ. GS acknowledges the projects PRIN-MUR 2020 MUR BEYOND-2p (``Astrochemistry beyond the second period elements'', Prot. 2020AFB3FX) and INAF-Minigrant 2023 TRIESTE (``TRacing the chemIcal hEritage of our originS: from proTostars to planEts''). The authors acknowledge the \textit{Kultrun Astronomy Hybrid Cluster} for providing HPC resources that have contributed to the research results reported in this paper.
\end{acknowledgements}

 \bibliographystyle{aa}
 \bibliography{Literature}

\onecolumn

\begin{appendix} 

\section{Derivation of \cite{Bovino20} (BFL20) \label{app:Ste}}
We now follow the derivation of Eq.~\eqref{eq:crir_observables}. The main reactions in our framework, considering the different isomers and isotopologues (but $\rm D_3^+$) for the formation of HCO$^+$ and DCO$^+$ are:

\begin{eqnarray*}
\ohdp + \mathrm{CO} \xrightarrow{{k_1}} \rm DCO^+ + oH_2\\
\phdp + \mathrm{CO} \xrightarrow{{k_2}}\rm  DCO^+ + pH_2\\
\pdhp + \mathrm{CO} \xrightarrow{{k_3}}\rm  DCO^+ + HD\\
\odhp + \mathrm{CO} \xrightarrow{{k_4}}\rm  DCO^+ + HD\\
\hp + \mathrm{CO} \xrightarrow{{k_5}}\rm   HCO^+ + pH_2\\
\hp + \mathrm{CO} \xrightarrow{{k_6}} \rm  HCO^+ + oH_2\\
\ho + \mathrm{CO} \xrightarrow{{k_7}} \rm  HCO^+ + oH_2\\
\pdhp + \mathrm{CO} \xrightarrow{{k_8}}\rm  HCO^+ + pD_2\\
\hp + \mathrm{CO} \xrightarrow{{k_6}} \rm  HCO^+ + oH_2\\
\ho + \mathrm{CO} \xrightarrow{{k_7}} \rm  HCO^+ + oH_2\\
\pdhp + \mathrm{CO} \xrightarrow{{k_8}}\rm  HCO^+ + pD_2\\
\odhp + \mathrm{CO} \xrightarrow{{k_9}}\rm  HCO^+ + oD_2\\
\phdp + \mathrm{CO} \xrightarrow{{k_{10}}}\rm  HCO^+ + HD\\
\ohdp + \mathrm{CO} \xrightarrow{{k_{11}}}\rm  HCO^+ + HD\\
\end{eqnarray*}

\noindent For the destruction, we consider only dissociative recombinations

\begin{eqnarray*}
    \mathrm{DCO^+ + e^-} \xrightarrow{\beta_1} \mathrm{D + CO}\\
    \mathrm{HCO^+ + e^-} \xrightarrow{\beta_2} \mathrm{H + CO}
\end{eqnarray*}

\noindent The kinetic equations read as
\begin{equation*}
    \frac{dn(\mathrm{DCO^+})}{dt} = k_1 n(\mathrm{CO}) n(\ohdp) + k_2 n(\mathrm{CO}) n(\phdp) + k_3 n(\mathrm{CO}) n(\pdhp) + k_4 n(\mathrm{CO}) n(\odhp) - \beta_1 n(\mathrm{DCO^+}) n(\mathrm{e^-}) 
\end{equation*}
and
\begin{equation*}
    \begin{split}
    \frac{dn(\mathrm{HCO^+})}{dt} = & \; (k_5+k_6) n(\mathrm{CO}) n(\hp) + k_7 n(\mathrm{CO}) n(\ho) + k_8 n(\mathrm{CO}) n(\pdhp) + k_9 n(\mathrm{CO}) n(\odhp)\; + \\
    & + k_{10} n(\mathrm{CO}) n(\phdp)+ k_{11} n(\mathrm{CO}) n(\ohdp)- \beta_2 n(\mathrm{HCO^+}) n(\mathrm{e^-}) \, .
    \end{split}
\end{equation*}

\noindent 
Assuming steady-state\footnote{We note that this is not the global steady-state of the system, but rather a local balance between formation and destruction at a given time which is keeping the DCO$^+$ and HCO$^+$ abundances constant.} and taking the ratio between the two equations we obtain

\begin{equation*}
    \begin{split}
    R_D & = \frac{n(\mathrm{DCO^+})}{n(\mathrm{HCO^+})} =  \left(\frac{\beta_2}{\beta_1}\right) \times \\
    & \frac{k_1 n(\mathrm{CO}) n(\ohdp) + k_2 n(\mathrm{CO}) n(\phdp) + k_3 n(\mathrm{CO}) n(\pdhp) + k_4 n(\mathrm{CO}) n(\odhp)}{(k_5+k_6) n(\mathrm{CO}) n(\hp) + k_7 n(\mathrm{CO}) n(\ho) + k_8 n(\mathrm{CO}) n(\pdhp) + k_9 n(\mathrm{CO}) n(\odhp)  + k_{10} n(\mathrm{CO}) n(\phdp)+ k_{11} n(\mathrm{CO}) n(\ohdp)} \; .
    \end{split}
\end{equation*}

\noindent Using the following relations among the reaction rates
\begin{itemize}
\item[] $k_1 = k_2$ \; ,
\item[] $k_3 = k_4$\; ,
\item[] $k_5+k_6 = k_7$\; ,
\item[] $k_8 = k_9$\; ,
\item[] $k_{10} = k_{11}$\; ,
\item[] $\beta_2=\beta_1$\; ,
\item[] 
\end{itemize}

\noindent the final equation reads (note that $n(\mathrm{CO})$ and $n(\mathrm{e^-})$ cancel out)
\begin{equation}
    R_D = \frac{n(\mathrm{DCO^+})}{n(\mathrm{HCO^+})} = \frac{k_1 \left [n(\ohdp) + n(\phdp) \right ]  + k_3 \left [n(\pdhp) + n(\odhp) \right ] }{k_7  \left [n(\ho) + n(\hp) \right ] + k_8 \left [n(\pdhp) + n(\odhp)\right ] + k_{10} \left [n(\phdp)+n(\ohdp) \right ]}\; . \label{eq:RD_bovino_tot}
\end{equation}
In order to simplify Eq.~\eqref{eq:RD_bovino_tot}, we can exploit further relations between the reaction rates, mainly linked to their branching ratios: $k_1/k_7=1/3$, $k_1/k_{10} = 1/2$, $k_8/k_3 = 1/2$, and $k_3/k_7=2/3$. Moreover, we neglect the correction for para and ortho species that cannot be observed, the contribution from the doubly deuterated isotopologue, {and the formation channel of \hcop via \hhdp}, arriving at\footnote{ This equation is accurate for small deuteration fraction ($R_\mathrm{D} \lesssim 10$\%). Above this level, the correction $1-2 R_\mathrm{D}$ needs to be taken into account, due to the formation of \hcop via \hhdp (cf. reaction rates $k_{10}$ and $k_{11}$).} 
\begin{equation}
    n(\mathrm{H_3^+}) = \frac{1}{3} \frac{n(\mathrm{oH_2D^+})}{R_D} \, . 
    \label{eq:Rd_h3p}
\end{equation}\\

\noindent 
\par
By inserting Eq.~\eqref{eq:Rd_h3p} in \citep[see e.g.][and also the derivation in Appendix~\ref{app:Paola}]{Oka19}
\begin{equation}
\zeta_2 = \frac{k_7 n(\mathrm{H_3^+}) n(\mathrm{CO})}{n(\mathrm{H_2})} \; , 
\label{eq:Oka}
\end{equation}



\noindent we obtain the final formula for the cosmic-ray ionisation rate
\begin{equation}
\zeta_2 = \frac{1}{3}\frac{k_7 n(\mathrm{oH_2D^+})}{R_D} \frac{n(\mathrm{CO})}{n(\mathrm{H_2})} \; .
\label{eq:CRIR_Bovino}
\end{equation}
This equation is valid as long as \ohhdp is the dominant deuterated form of H$_3^+$ and when the reaction with CO is more important than dissociative recombination in the destruction of $\rm H_3^+$. The first limitation implies that when deuteration levels become higher and \ohhdp is converted into its doubly and triply deuterated isotopologues, Eq.~\eqref{eq:CRIR_Bovino} cannot be used anymore. Concerning the destruction pathways, we can investigate at which CO abundance (as a function of the electronic fraction) its reaction with $\rm H_3^+$ dominates over the dissociative recombination (see the right-hand side of Eq.~\eqref{eq:h3p_kinetics}). We find that this holds for $f_\mathrm{D} < 9\times 10^{-7}/x(\mathrm{e^-})$ (assuming $f_\mathrm{para} = 0.7$, see also Appendix~\ref{app:Paola}). For $x(\mathrm{e}) = 10^{-8}$, close to the values found in the reference run, the reaction with CO is dominant if $f_\mathrm{D}\lesssim 90$, which is verified in our simulations. However, at very high densities, when $f_\mathrm{D}\gtrsim 100$, this assumption might not hold anymore. 
\par
By introducing average quantities integrated over the path $L$ along the line of sight, Eq.~\eqref{eq:CRIR_Bovino} finally becomes

  \begin{equation}
    \zeta_2 = k_7 \frac{1}{3}\times X(\mathrm{CO}) \times \frac{N_\mathrm{col}(\mathrm{oH_2D^+})}{R_\mathrm{D}}\frac{1}{L}    \; .
    \label{eq:crir_observables_app}
    \end{equation}
Note that $k_7 = k_\mathrm{CO}^{\rm oH_3^+}$ in the main text.
    


\section{Derivation of \cite{Caselli98} (CWT98) \label{app:Paola}}
We now illustrate the derivation of the equations used in \cite{Caselli98}, which in turn are based on previous works (e.g. \citealt{Guelin77, Guelin82}). In particular, we aim to follow the same approach as those papers, including this time the ortho/para separation for all the involved species. \par
The first part of the equations is the same as illustrated in Appendix~\ref{app:Ste}, and involves balancing the destruction and formation pathways of \hcop and \dcop, arriving at (see Eq.~\eqref{eq:RD_bovino_tot})

\begin{equation}
R_D = \frac{1}{3} \frac{n(\mathrm{H_2D^+})}{n(\rm H_3^+)} \; ,
\label{eq:Rd_Caselli}
\end{equation}
 where we neglect the reactions involving doubly and triply deuterated H$_3^+$ and reactions 10 and 11). In this case, however, we look for a way to express the ratio $n(\mathrm{H_2D^+}) / n(\rm H_3^+)$ as a function of the densities of CO, HD, and of the electron fraction. To this aim, we have to write the reactions involved in the formation and destruction of \hhdp (in its ortho and para states). For the formation pathways, we have

\begin{eqnarray*}
\mathrm{ pH_3^+ + HD} \xrightarrow{{k_{12}}} \rm \phdp + pH_2 \\
\mathrm{ pH_3^+ + HD} \xrightarrow{{k_{13}}} \rm \phdp + oH_2 \\
\mathrm{ pH_3^+ + HD} \xrightarrow{{k_{14}}} \rm \ohdp + pH_2 \\
\mathrm{pH_3^+ + HD} \xrightarrow{{k_{15}}}  \rm\ohdp + oH_2 \\
\mathrm{ oH_3^+ + HD} \xrightarrow{{k_{16}}} \rm \phdp + oH_2 \\
\mathrm{ oH_3^+ + HD} \xrightarrow{{k_{17}}} \rm \ohdp + oH_2 \\
\mathrm{ oH_3^+ + HD} \xrightarrow{{k_{18}}} \rm \ohdp + pH_2
\end{eqnarray*}
The destruction pathways, instead, involve the reactions with CO (with rates $k_1$, $k_2$, $k_{10}$, and $k_{11}$, see above)\footnote{Note that CWT98 assumed an equal abundance of atomic oxygen O as of CO, and assumed also the same destruction rates, to add these pathways to the final equations.}, and the following dissociative recombinations
\begin{equation*}
\begin{split}
\rm \ohdp + e^- & \xrightarrow{{\beta_3}}\rm H + H + D\\
\rm \ohdp + e^- &\xrightarrow{{\beta_4}} \rm D + oH_2\\
\rm \ohdp + e^- &\xrightarrow{{\beta_5}}\rm H + H\\
\rm \phdp + e^- & \xrightarrow{{\beta_6}}\rm  H + H + D\\
\rm \phdp + e^- & \xrightarrow{{\beta_7}}\rm  D + pH_2\\
\rm \phdp + e^- & \xrightarrow{{\beta_8}}\rm  H + HD
\end{split}
\end{equation*}
Note that we have neglected all the reactions involving doubly and triply deuterated $\rm H_3^+$, to be consistent with the simplification done to obtain Eq.~\eqref{eq:Rd_Caselli}. For several of the involved reaction rates, it is possible to show that
\begin{itemize}
\item[] $k_{12} + k_{13} + k_{14} +k_{15} \approx k_{16} + k_{17} + k_{18} $
\item[] $\beta_3 \approx \beta_6$
\item[] $\beta_4 \approx \beta_7$
\item[] $\beta_5 \approx \beta_8$
\end{itemize}
These relations do not hold exactly, but we will show that in the temperature range here considered the agreement is reasonably good. The first relation is reported in the left panel of Fig.~\ref{fig:rates}. For temperatures $\lesssim 25\, \rm K$, the discrepancy is lower than 10\%, and in the range $10-15\, \rm K$, the difference is $1-3$\%. We hence assume that equality holds. For the various rates of dissociative recombination ($\beta_3$ to $\beta_8$), the difference at $15\, \rm K$ is $\approx 6$\%, but it quickly rises above 25\% outside the range $10-20 \, \rm K$. We hence suggest extreme caution in using these and the following relations outside this temperature range. However, without these assumptions, it is in practice impossible to properly re-derive and upgrade the formula proposed by CWT98.

\begin{figure}
    \centering
    \includegraphics[width=.85\textwidth]{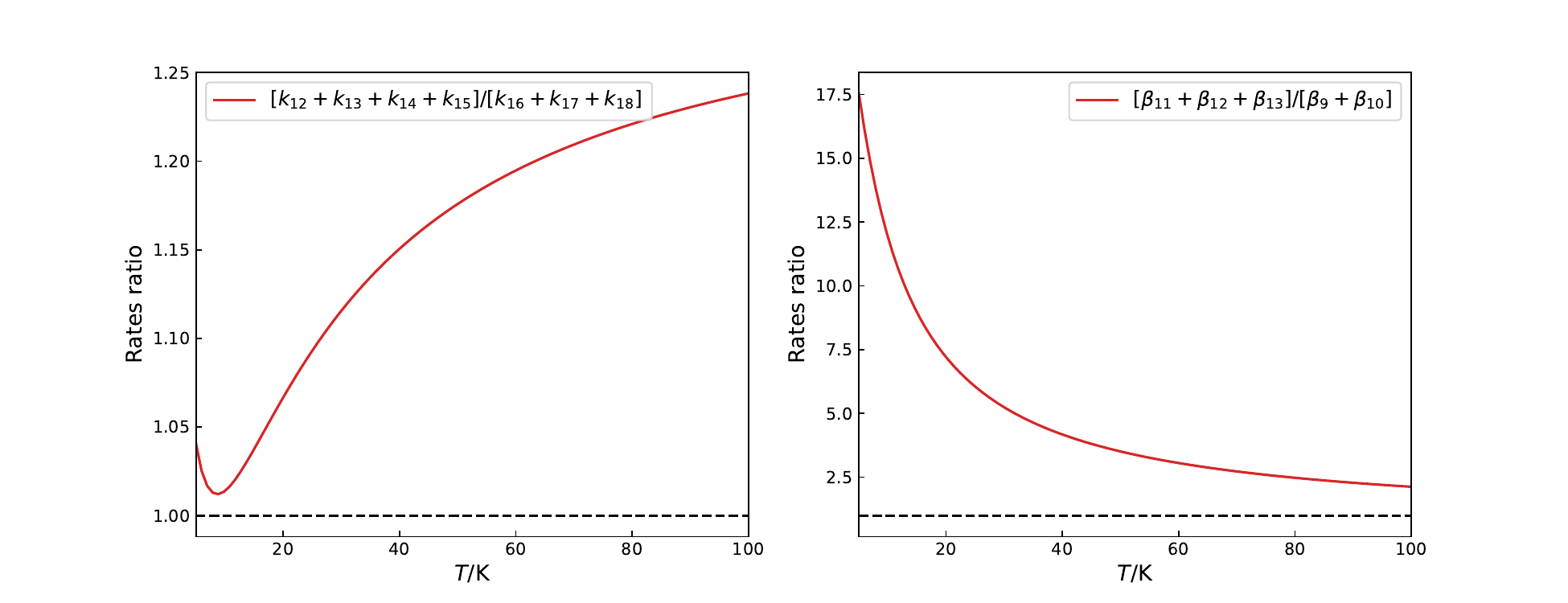}
    \caption{Left panel: the ratio between the sum of the rates of the reactions between $\rm pH_3^+$ and HD and those involving  $\rm oH_3^+$ and HD, as a function of temperature. Right panel: the ratio between the total rate of dissociative recombination of $\rm pH_3^+$ {and} that of $\rm oH_3^+$, as a function of temperature. In both panels, the horizontal dashed lines correspond to unity.}
    \label{fig:rates}
\end{figure}

We can now write the kinetic equations for the para and ortho species separately as

\begin{equation*}
 \begin{split}
  \frac{dn(\ohdp)}{dt} & =  {k}_{14} n(\mathrm{pH_3^+}) n(\mathrm{HD}) + {k}_{15} n(\mathrm{pH_3^+}) n(\mathrm{HD}) + {k}_{17} n(\mathrm{oH_3^+}) n(\mathrm{HD})+ {k}_{18} n(\mathrm{oH_3^+}) n(\mathrm{HD}) \\ & - \mathrm{k}_{1} n(\mathrm{\ohdp}) n(\mathrm{CO})  - {k}_{11} n(\mathrm{\ohdp}) n(\mathrm{CO})  -  \beta_3 n(\ohdp) n(\mathrm{e^-}) - \beta_4 n(\ohdp) n(\mathrm{e^-}) -\beta_5 n(\ohdp) n(\mathrm{e^-})    \; , 
 \end{split}   
\end{equation*}
and
\begin{equation*}
\begin{split}
 \frac{dn(\phdp)}{dt} & =  {k}_{12} n(\mathrm{pH_3^+}) n(\mathrm{HD}) + {k}_{13} n(\mathrm{pH_3^+}) n(\mathrm{HD}) + {k}_{16} n(\mathrm{oH_3^+}) n(\mathrm{HD}) - {k}_{2} n(\mathrm{\phdp}) n(\mathrm{CO}) +\\ &
- {k}_{10} n(\mathrm{\phdp}) n(\mathrm{CO})  -  \beta_6 n(\phdp) n(\mathrm{e^-}) - \beta_7 n(\phdp) n(\mathrm{e}) -\beta_8 n(\phdp) n(\mathrm{e^-})     \; .
\end{split}
\end{equation*}
Assuming the steady state, we can re-write the two equations above as
\begin{equation*}
\begin{split}
({k}_{14}+  {k}_{15} )n(\mathrm{pH_3^+}) n(\mathrm{HD}) +   ({k}_{17}+ {k}_{18} )n(\mathrm{oH_3^+}) n(\mathrm{HD}) & = ({k}_{1}+{k}_{11}) n(\mathrm{\ohdp}) n(\mathrm{CO})  +( \beta_3+ \beta_4 + \beta_5) n(\ohdp) n(\mathrm{e^-}) \; , \\
({k}_{12}+  {k}_{13} )n(\mathrm{pH_3^+}) n(\mathrm{HD})+ {k}_{16} n(\mathrm{oH_3^+}) n(\mathrm{HD}) & = ({k}_{2}+{k}_{10}) n(\mathrm{\phdp}) n(\mathrm{CO}) + (\beta_6 +\beta_7+ \beta_8 )n(\phdp) n(\mathrm{e^-}) \; .
\end{split}
\end{equation*}
By summing the two equations above and exploiting the relations between the reaction rates, we arrive at:
\begin{equation*}
({k}_{12}+  {k}_{13}+  {k}_{14}+ {k}_{15} )n(\mathrm{H_3^+}) n(\mathrm{HD}) = ({k}_{1}+{k}_{11}) n(\mathrm{H_2D^+}) n(\mathrm{CO})  +( \beta_3+ \beta_4 + \beta_5) n(\mathrm{H_2D^+}) n(\mathrm{e^-})   \; ,  
\end{equation*}
which allows us to rewrite Eq.~\eqref{eq:Rd_Caselli} as
\begin{equation}
R_\mathrm{D} = \frac{1}{3} \frac{n(\rm{H_2D^+})}{n(\rm H_3^+)} = \frac{1}{3} \frac{({k}_{12}+  {k}_{13}+  {k}_{14}+ {k}_{15} ) n(\mathrm{HD})}{({k}_{1}+{k}_{11}) n(\mathrm{CO})  +( \beta_3+ \beta_4 + \beta_5) n(\mathrm{e^-})  } \; .
\label{eq:RD_newrates}
\end{equation}

The next step is to express the quantity $R_\mathrm{H} = n(\mathrm{HCO^+})/n(\mathrm{CO})$ as a function of the cosmic ray ionisation rate and the electronic fraction. First, we solve the kinetic equation for \hcop in steady-state, again neglecting all terms containing $\rm D_2H^+$ and D$_3^+$ (see above):
\begin{equation}
R_\mathrm{H} = \frac{n(\mathrm{HCO^+})}{n(\mathrm{CO} )} =  \frac{k_7 n(\mathrm{H_3^+})}{\beta_2 n(\mathrm{e^-})}  \; .
\label{eq:RH_newrates}
\end{equation}
To find an expression for $n(\mathrm{H_3^+})$, we solve its kinetics. Its total formation rate is $\zeta_2 n(\mathrm{H_2})$. The destruction pathways instead are separated in the ortho and para species, and involve the reaction with CO (reactions 5 to 7) and the following dissociative recombinations:
\begin{equation*}
\begin{split}
\rm oH_3^+ + e^- &\xrightarrow{\beta_9} \rm H + H + H\\
\rm oH_3^+ + e^- &\xrightarrow{\beta_{10}} \rm H + oH_2\\
\rm pH_3^+ + e^- & \xrightarrow{\beta_{11}} \rm  H + H + H\\
\rm pH_3^+ + e^- & \xrightarrow{\beta_{12}}\rm  H + pH_2\\
\rm pH_3^+ + e^- & \xrightarrow{\beta_{13}} \rm H + oH_2
\end{split}
\end{equation*}
The destruction rates of the two species are then:
\begin{equation*}
\begin{split}
  \left . \frac{dn( \mathrm{oH_3^+})}{dt} \right |_\mathrm{destr} & = - \left [ k_7 n(\mathrm{oH_3^+}) n(\mathrm{CO})+ \left ( \beta_9 + \beta_{10} \right ) (\mathrm{oH_3^+})  n(\mathrm{e^-}) \right ]  \; ,\\
 \left . \frac{dn( \mathrm{pH_3^+})}{dt} \right |_{\mathrm{destr}} &= - \left [ \left( k_5 + k_6 \right )n(\mathrm{pH_3^+}) n(\mathrm{CO})+ \left ( \beta_{11} + \beta_{12} +\beta_{13} \right ) (\mathrm{pH_3^+})  n(\mathrm{e^-}) \right ]   \; .
\end{split}
\end{equation*}
From these equations, we can then compute the destruction rate for the total $\rm H_3^+$ density, and set it equal to the total formation rate, obtaining:
\begin{equation}
\zeta_2 n(\mathrm{H_2}) = k_7 n(\mathrm{H_3^+}) n(\mathrm{CO}) + n(\mathrm{e^-}) \left[ \left( \beta_{11} + \beta_{12} +\beta_{13} \right ) n(\mathrm{pH_3^+}) + \left ( \beta_9 + \beta_{10} \right ) (\mathrm{oH_3^+})  \right ]  \; .
\label{eq:h3p_kinetics_full}
\end{equation}

To further simplify Eq.~\eqref{eq:h3p_kinetics_full}, we focus on the dissociative recombination rates of ortho- and para-H$_3^+$. Their ratio is shown in the right panel of Fig.~\ref{fig:rates}, where one can see how at low temperatures ($T\lesssim 20\, \rm K$), the reaction rates of $\rm pH_3^+$ is about one order of magnitude higher than that of $\rm oH_3^+$. We will hence neglect the second term, and introduce the para fraction $f_\mathrm{para} = \rm pH_3^+/ H_3^+$, to write:
\begin{equation}
\zeta_2 n(\mathrm{H_2}) = k_7 n(\mathrm{H_3^+}) n(\mathrm{CO}) + n(\mathrm{e^-}) \beta_\mathrm{pH_3^+}   n(\mathrm{H_3^+}) f_\mathrm{para}     \; , 
\label{eq:h3p_kinetics}
\end{equation}
where $\beta_\mathrm{pH_3^+}  = \beta_{11} + \beta_{12} +\beta_{13}$. Equation~\eqref{eq:h3p_kinetics} can be solved for $n(\rm H_3^+)$, and then inserted in Eq.~\eqref{eq:RH_newrates}. The system of equations to infer the electron fraction and \crir becomes:
\begin{equation}
\begin{split}
x(\mathrm{e}^-) & =\frac{1}{ \beta_3+ \beta_4 + \beta_5} \left [ \frac{1}{3} \frac{(\mathrm{k}_{12}+  \mathrm{k}_{13}+  \mathrm{k}_{14}+ \mathrm{k}_{15} ) X (\mathrm{HD})}{R_\mathrm{D}} - (\mathrm{k}_{1}+\mathrm{k}_{11}) X ( \mathrm{CO} ) \right ] \; , \\
\zeta_2 & = \left [ k_7  X(\mathrm{CO}) + \beta_\mathrm{pH_3^+}  \, f_\mathrm{para} \,  x(\mathrm{e^-})   \right ] \frac{ \beta_2  R_\mathrm{H} x(\mathrm{e^-})   n(\mathrm{H_2}) } {k_7} \; ,   
\label{eq:PaolaNew}
\end{split}
\end{equation}
where now quantities are expressed in terms of abundances, rather than volume densities.
\par

Equations~\eqref{eq:PaolaNew} have a mathematical limitation, in that for certain combinations of $R_\mathrm{D}$ and $X\rm (CO)$ (or, equivalently, of $f_\mathrm{D}$) the first one yields negative values for the electron fraction. By computing the reaction rates at $15\, \rm K$, and assuming $ X (\mathrm{HD}) = 1.5 \times 10^{-5}$ and $f_\mathrm{para} = 0.7$,  we find that the threshold is $R_\mathrm{D} = 0.029 \times f_\mathrm{D}$. {Note a small variation to the original limitation of $R_\mathrm{D} = 0.023 \times f_\mathrm{D}$.} For deuteration levels higher than this limit, the equation cannot be applied. At $10\, \rm K$, the new rates of Eqs.~\eqref{eq:PaolaNew} $10-50\%$ are lower than the original equations (3 and 4) of CWT98. As a result, the updated equations lead to electron fractions lower by 20\% and \crir values lower by 50\% than the original derivation.
\par
Figure~\ref{fig:gelato} shows the dependency of \crir (normalised by the quantity $R_\mathrm{H}\times n \rm (H_2)$) as a function of the deuterium fraction and CO depletion factor. One can see that, for decreasing deuteration levels, the quantity $\zeta_2/(R_\mathrm{H}\times n \rm (H_2))$ increases by several orders of magnitude. Since \crir depends linearly on $R_\mathrm{H}$ and $n \rm (H_2)$, this translates into an equal increase also of this quantity. The plot also shows that for the deuteration values observed in dense prestellar cores ($R_\mathrm{D} = 0.01-0.1$), there is a strong dependency on the depletion factor, up to $f_\mathrm{D} \approx 10-20$.
\begin{figure}[!h]
    \centering
    \includegraphics[width=0.5 \textwidth]{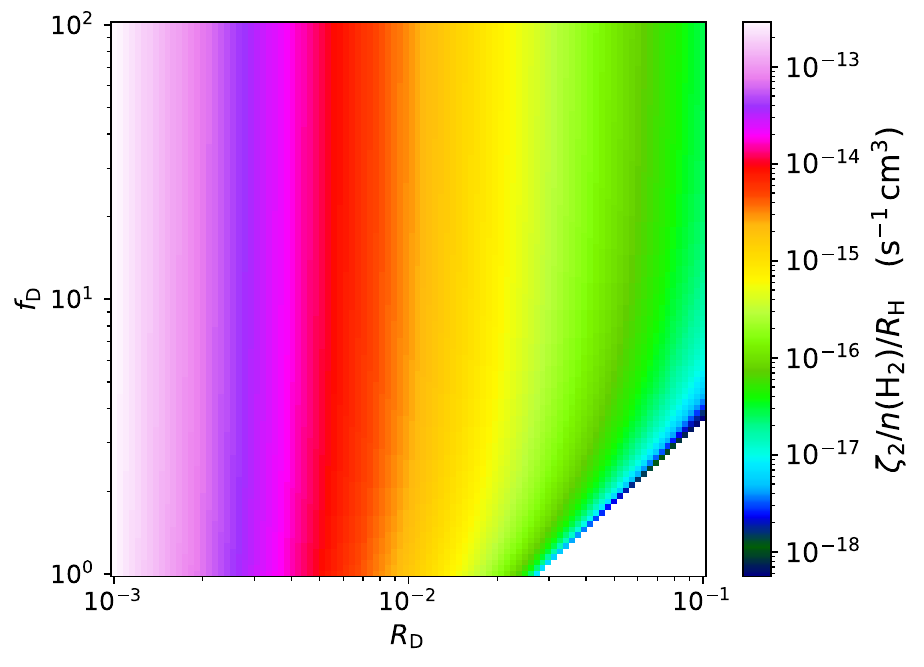}
    \caption{\crir values (divided by $R_\mathrm{H}\times n \rm (H_2)$) obtained with Eq.~\eqref{eq:PaolaNew} at $15\, \rm K $, as a function of $R_\mathrm{D}$ and $f_\mathrm{D}$. The bottom-right corner is missing because it violates the condition  $R_\mathrm{D} < 0.029 \times f_\mathrm{D}$. The plot shows that the quantity $\text{\crir} /(R_\mathrm{H}\times n \rm (H_2))$ increases by up to four orders of magnitude when the deuteration fraction decreases from $0.1$ to $10^{-3}$ and that the CO depletion factor plays a significant role at $R_\mathrm{D}$ typical for dense gas ($0.01-0.1$).}
    \label{fig:gelato}
\end{figure}

\section{\tex values for \dcop and \htcop in LVG analysis\label{app:tex}}

In Sect.~\ref{sec:col_density} we discussed the choice of \tex values for each transition processed with the LVG radiative transfer. Whilst those for \cdo and \ohhdp are well documented in the literature, this is not the case for \dcop and \htcop transitions. We have corroborated the values we chose by using the online tool \textsc{radex}. However, in this Appendix, we investigate how a variation of the \tex values of $\approx$20\% affect the inferred \crir values. The results are shown in Fig.~\ref{fig:tex_variation}, where, in the upper panel, we analyse run~5 that uses the \dcop and \htcop (3-2) lines, and in the lower panel we present run~6, which instead focuses on the first rotational transitions.
\par
In run 5, we explore a \tex variation of $1\,$K. The resulting \crir values change by $\approx 25\%$. Note that the variation is stronger when the \tex is lowered, due to the \tex dependence of Eq.~\ref{Ncol_tau}. For run 6, the \text values are changed by 2$\,$K, exploring the range $8-12\,$K, which leads to a smaller variation of the inferred \crir maps (less than 10\%). In all cases, the derived \crir values do not change significantly {compared to} the uncertainties, and the median values are still less {than} a factor of {three} from the actual input \crirs of the simulation.

\begin{figure}[!h]
    \centering
    \includegraphics[width=0.7 \textwidth]{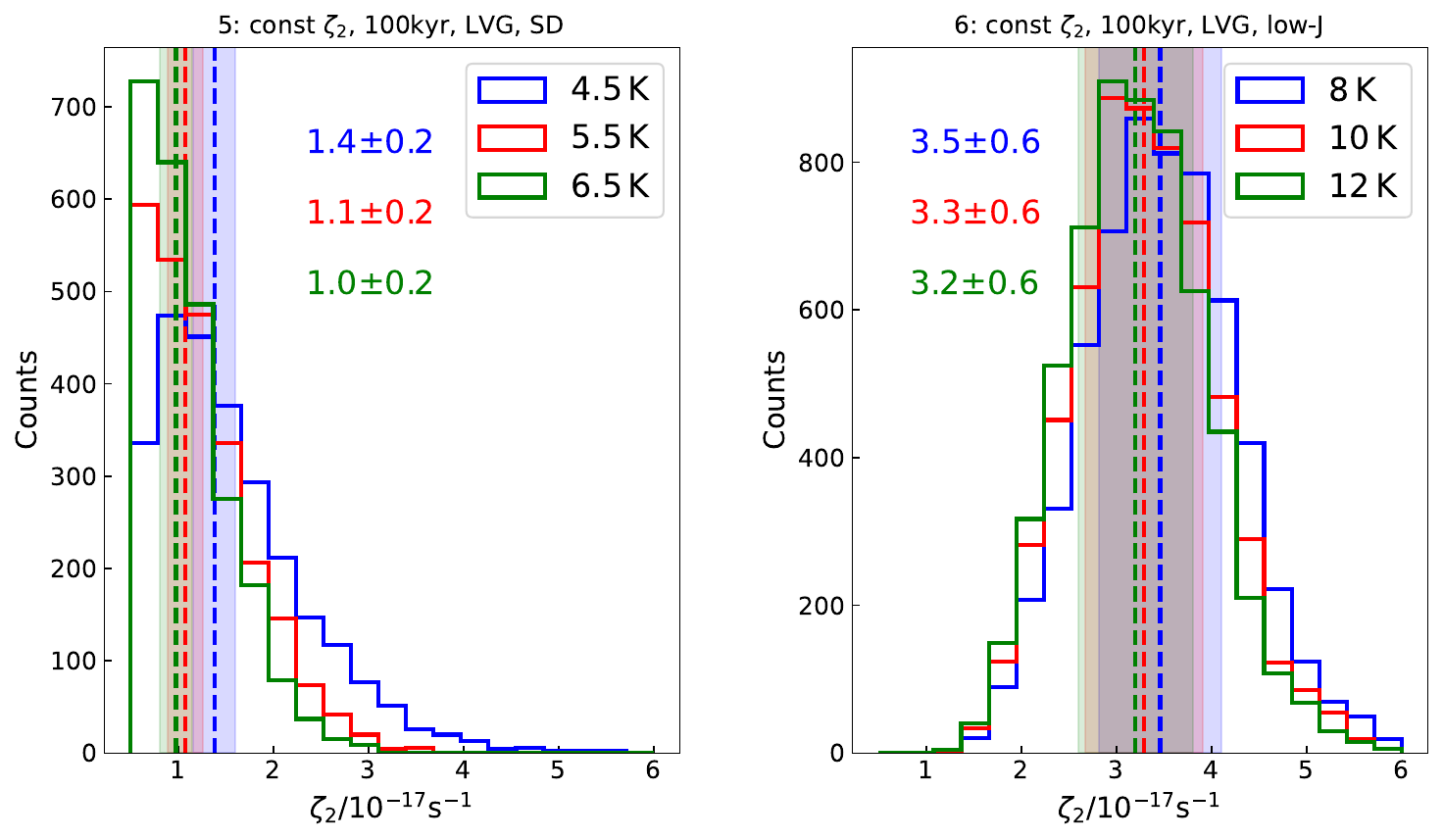}
    \caption{Histograms of the derived \crir values with the BFL20 method in run 5 (upper panel, high-J transitions) and run 6 (lower panel, low-J transitions). The different colours show different assumptions for the excitation temperature of \dcop and \htcop (assumed to be equal). In particular, the reference values used in the main text are shown in red (hence these data are the same presented in Fig.~\ref{fig:crir_histograms}), whilst the blue/green curves show a positive/negative variation of $\approx 20$\% of that value, respectively (labelled in the top-right corner of each panel). The median values and uncertainties are shown with the vertical dashed lines and shaded areas and with coloured text in each panel. 
    \label{fig:tex_variation}}
\end{figure}
\section{The choice of $L$ in single-dish-like runs\label{app:L_study}}

 To apply the analytical equations in the case of single-dish-like post-processing of the simulations, we set $L=0.3\, \rm pc$, corresponding to the size of the analysed cut-out from the original simulation box ($\rm 0.6 \times 0.6 \times 0.6 \rm \, pc^3$). To support this choice and to show that this is the relevant physical quantity to take into account, we perform a further additional test. We analyse the same simulation as in the reference run n.~2, but this time we halve the size of the cut-out box before the radiative transfer. Hence, $L=0.15\, \rm pc$. The subsequent steps in terms of radiative transfer and post-processing are identical to those followed in run 2. \par
 Our simulation consists of a box filled with molecular gas. As a consequence, most of the analysed quantities change when $L=0.15\, \rm pc$. The \ohhdp column density is the only one that is not affected by this change since this molecule is highly concentrated in the densest part of the core and has a high critical density. \htcop and \cdo suffer the largest change, and their retrieved column densities decrease by a factor of up to $2-3$. In the simulation, these two molecules are abundant everywhere, and hence cutting a smaller portion of the simulation box affects significantly their total density on the line of sight. The \dcop column density decreases by $10-20$\%. The total gas density $N(\rm H_2)$ derived from the dust thermal emission decreases marginally ($\sim ~5$\% or less). Because of these changes, $R_\mathrm{D}$ increases and $X\rm(CO)$ decreases, however not at the same rate. We compare the \crir maps obtained in the two tests with distinct $L$ computed with the BFL20 method in Fig.~\ref{fig:L_study} (left and middle panel). The distributions of values are shown as histograms in the right panel. The maps are similar both in morphology and in the range of values. The histograms confirm these conclusions. The distributions are comparable and the median values (shown with vertical, dashed lines) are consistent with each other{: ${\langle \zeta_2 \rangle = (4.0 \pm 0.7 ) \times 10^{-17} \, \rm s^{-1}}$ (${L=0.3\, \rm pc}$) and ${\langle \zeta_2 \rangle = ( 3.7\pm 0.8 ) \times 10^{-17} \, \rm s^{-1}}$ (${L=0.15\, \rm pc}$)}. If we used the old value $L=0.3 \rm \, pc$ in the new test with a smaller box, the blue histogram would shift to the left (towards lower values) by a factor of two.
\par
This test suggests that using the size of the cut-out box to estimate $L$ in these runs is an appropriate choice. We stress again that this is a consequence of the particular simulation we are investigating, which represents a rather dense medium where most of the molecules of interest are abundant in the entire box. This is not the case for isolated cores such as the ones tested in Sect.~\ref{sec:realObs} of the manuscript, for which the prescription based on the $N(\rm H_2)$ isocontour is appropriate.

\begin{figure}[!h]
    \centering
    \includegraphics[width=\textwidth]{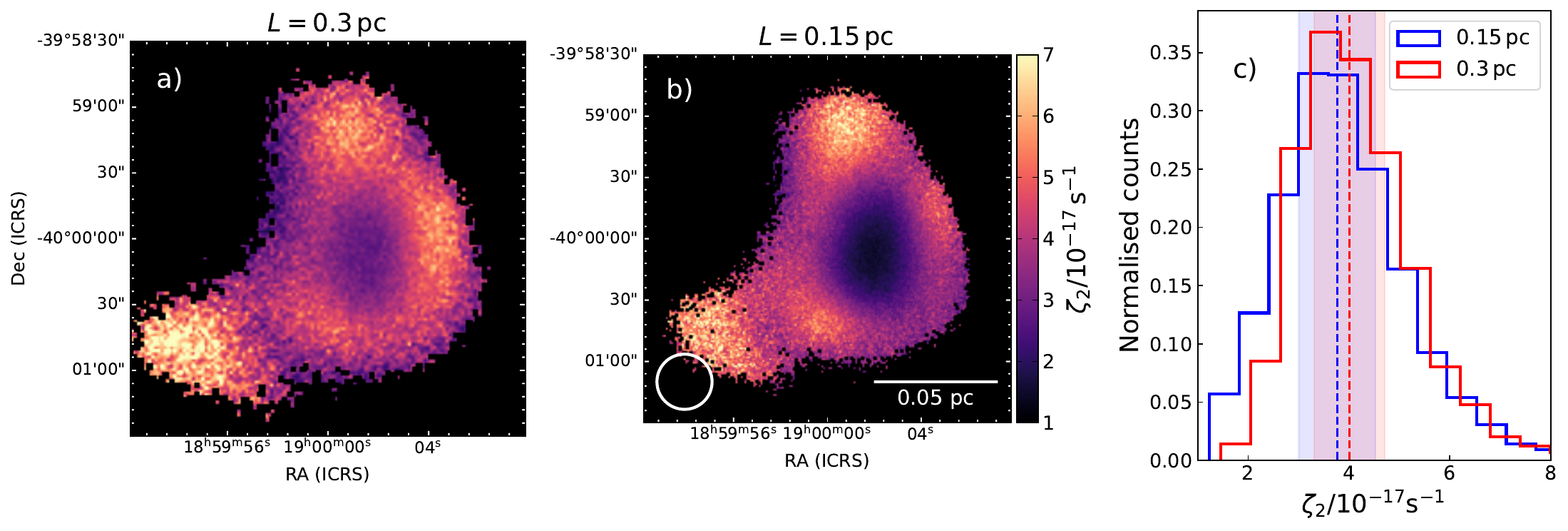}
    \caption{Comparison of the resulting \crir maps (in units of $10^{-17}\, \rm s^{-1}$) on the two runs with distinct box sizes $L$ but otherwise identical, using the BFL20 method. Panel a) shows the run with $L=0.3\, \rm pc$ (i.e. the reference run 2), and panel b) shows the results with $L=0.15 \rm \, pc$. The colorbar is kept fixed to ease the comparison. The histogram distributions are compared in panel c), labelled in the top-right corner. The median values (median uncertainties) are shown with the vertical dashed lines (shaded areas).}
    \label{fig:L_study}
\end{figure}

\end{appendix}
%
%

\end{document}